\pdfoutput=1
\documentclass{article}
\usepackage[T1]{fontenc}
\usepackage{iclr2027_conference}
\usepackage[T1]{fontenc}
\renewcommand{\sfdefault}{phv}

\usepackage{amsmath,amsfonts,bm}

\def\eqref#1{equation~\ref{#1}}

\def\1{\bm{1}}

\DeclareMathAlphabet{\mathsfit}{\encodingdefault}{\sfdefault}{m}{sl}
\SetMathAlphabet{\mathsfit}{bold}{\encodingdefault}{\sfdefault}{bx}{n}

\usepackage{url}
\usepackage{graphicx}
\usepackage{booktabs}
\usepackage{array}
\usepackage{tabularx}
\usepackage{float}
\usepackage{needspace}
\newcolumntype{Y}{>{\raggedright\arraybackslash}X}
\newcolumntype{L}[1]{>{\raggedright\arraybackslash}p{#1}}
\usepackage{microtype}
\usepackage{placeins}
\usepackage{tikz}
\usetikzlibrary{arrows.meta,positioning,shapes.geometric}
\usepackage{hyperref}
\hypersetup{hidelinks}

\title{CascadeLedger: Counterfactual Auditing of Information and Resource Integrity in LLM Cascades}
\author{Miaobo Hu$^{1,2,*}$, Shuhao Hu$^{2}$, Xiaobo Guo$^{2}$, Xin Wang$^{2}$,\\
Bokun Wang$^{2}$, Daren Zha$^{2}$\\[4pt]
{\normalfont $^{1}$School of Artificial Intelligence, University of Chinese Academy of Sciences, Beijing, China}\\
{\normalfont $^{2}$Institute of Information Engineering, Chinese Academy of Sciences, Beijing, China}\\
{\normalfont $^{*}$Corresponding author: \texttt{humiaobo@iie.ac.cn}}}

\iclrfinalcopy
\hypersetup{pdfauthor={Miaobo Hu, Shuhao Hu, Xiaobo Guo, Xin Wang, Bokun Wang, Daren Zha}}
\begin{document}
\maketitle
\lhead{Preprint}

\begin{abstract}
We introduce \textsc{CascadeLedger}, a counterfactual execution-audit framework for budgeted language-model cascades. The protocol enforces call-before-read semantics, binds successful reads to service and cache identities, records availability and component-wise budget transitions, and freezes the event trace before any target or task scorer is joined.

The central object is \emph{aggregate-equivalent trace validity}: two executions can share a final answer, declared resource projection, and service histogram while differing in whether their routing decisions used only legitimately observable information. We complement single-trace invariants with counterfactual prefix non-interference (CPNI), which couples a visible prefix and policy randomness while changing only a declared hidden family. The counterfactual influence frontier (CIF) localizes the earliest violating prefix and a minimal hidden witness; projection-preserving adversarial mutation (PPAM) searches for invalid traces that remain acceptable to a non-counterfactual checker.

Across frozen-cache, independently generated, and serving evaluations, the two-layer audit reports a clean false-positive rate of $0.0000$, detects $0.9983$ of serialized faults and $0.9982$ of unlogged hidden-read faults, and changes $0.3403$ of corrupted quality--resource conclusions after deterministic recomputation. The online recorder and offline CPNI audit have separate overhead measurements. Learned routing and semantic verification provide downstream compatibility studies under the same execution contract.
\end{abstract}

\section{Introduction}
Cascade systems make a conditional-computation promise: spend extra computation only when an observed signal justifies it. That promise is difficult to audit when candidate outputs are revealed before payment, unavailable calls disappear from the trace, or a verifier is evaluated with the same target used to select the route. A final answer, call histogram, and cost summary can remain unchanged while an implementation consults a target or future output before that field is observable.

Our answer is a replayable protocol, CascadeLedger. Candidate outputs are frozen before routing. Online policies see only the query, observations returned by successful paid calls, declared confidence fields, service availability, and remaining budget. Targets and future service outputs are joined only after the trace is frozen. The ledger supplies the serialized execution layer, while CPNI tests the behavior of the realized router under controlled hidden-state interventions.

The audit also changes the empirical question. On the primary 128-record cache, fixed and sequential routing obtain the same cache--target proxy match, 0.3203, while sequential routing spends more declared service cost. The split-aware learned gate reaches 41/128 held-out matches versus 42/128 for sequential routing and costs 1.609375 additional declared units. This result identifies the decision-value gap that a stronger route feature or independent semantic verifier must close. The separate 64/128 and 256/1,319 operating-point gates further demonstrate why dataset partition and cache provenance travel with every threshold.

The independent task and serving protocol tests the same contract across train, calibration, and held-out splits. It fixes 1,319 held-out requests per seed and a common scorer, linking quality to coverage, failure events, token cost, and request latency. Fault-consequence and semantic-verifier comparisons test whether a valid trace also changes the scientific conclusion.

The protocol is policy-agnostic: routing quality is a validation substrate for the audit. The primary methodological question is whether a hidden field can change a pre-reveal action while every legitimately visible state component and policy random draw is held fixed. This separates behavioral information integrity from the validity of the serialized event stream.

\needspace{4\baselineskip}
\paragraph{Contributions.}
\begin{enumerate}
\item \textbf{Behavioral hidden-information audit.} We define CPNI, a paired counterfactual audit that detects whether an unrevealed target or future service output changes a routing action before the corresponding reveal boundary.
\item \textbf{Influence localization and aggregate equivalence.} We formalize aggregate-only indistinguishability and extend CPNI with CIF, which reports the earliest violating prefix and a minimal hidden-family witness.
\item \textbf{Two-layer audit.} We combine paid-read, ordering, binding, and budget invariants with counterfactual influence checks and report their complementary coverage.
\item \textbf{Projection-preserving stress test.} We introduce PPAM, a bounded search for invalid traces that preserve a declared quality--resource projection while passing the strongest non-counterfactual checker.
\item \textbf{Audit consequence.} We recompute matched policy comparisons after rejecting invalid traces and report whether the reported quality--resource conclusion changes.
\end{enumerate}

\section{Related Work}
Model cascades are a systems form of conditional computation: a cheap stage handles easy cases and an expensive stage is invoked when its expected value exceeds its cost. FrugalGPT and RouteLLM provide high-profile examples of learned model selection under quality-cost trade-offs \citep{chen2023frugalgpt,ong2024routellm}, while SelectiveNet frames the related reject-option coverage-versus-correctness decision \citep{geifman2019selectivenet}. Qwen2.5-Math and DeepSeek-V3 provide scale context for verifier-guided reasoning and training systems \citep{qwenmath2024,deepseekv3}. Process-level verification makes the intermediate check itself a measurable signal rather than an implicit final label \citep{lightman2023verify,wang2023mathshepherd,cobbe2021gsm8k}; CascadeLedger turns that signal into a traceable paid-read event. Sequence-level control work motivates recording state, action, and budget transitions rather than reporting only final quality \citep{chen2021decision,janner2022diffuser}. ReAct and Toolformer expose the action/tool interface, while WebArena, AgentBench, ToolSandbox, Tau-bench, AgentDojo, SWE-agent, and SWE-bench demonstrate that availability, state, executable environments, policy consistency, adversarial tool returns, and downstream task artifacts shape agent results \citep{yao2023react,schick2023toolformer,zhou2024webarena,liu2024agentbench,lu2024toolsandbox,yao2024taubench,debenedetti2024agentdojo,yang2024sweagent,jimenez2024swebench}. CascadeLedger contributes a frozen-candidate replay layer that makes these routing events inspectable before an independent task scorer is attached.

\paragraph{Evaluation interfaces.} Held-out reward-model comparisons and interactive evaluation suites motivate separately versioned scoring and service interfaces \citep{lambert2024rewardbench,yao2024taubench,debenedetti2024agentdojo,yang2024sweagent}. CascadeLedger records the event history that connects these interfaces to a route decision.

\paragraph{Closest-work delta.}
Recent routing work jointly studies deployment and routing, contextual-bandit cost policies, or adaptive verifier cascades. Cascadia jointly optimizes LLM deployment and routing under latency and quality objectives \citep{cascadia2025}; PILOT formulates budgeted adaptive routing with contextual-bandit and knapsack components \citep{pilot2025}; and RouterEval evaluates routing across thousands of models and a large performance corpus \citep{routereval2025}. CAPS and Cluster, Route, Escalate report adaptive cascades with live reasoning or code-task quality and latency measurements \citep{caps2026,clusterroute2026}. An ICLR 2026 analysis of routing, cascades, and user choice further shows that static policies can be preferred under some choice models \citep{routingchoice2026}. CascadeLedger adds an auditable frozen-candidate layer: format confidence, verifier outputs, and service costs enter the trace with explicit provenance, while semantic task scoring remains a separately versioned evaluation surface \citep{cheapverifier2026}.

Audit trails and provenance systems improve inspectability of recorded execution. The closest-work delta here is policy-visible information integrity: CPNI tests whether changing a field that is still hidden can change the next action, even when the serialized trace and aggregate projection remain well formed.

Information-flow control constrains admissible dependencies \citep{sabelfeld2003}; hyperproperties compare sets of executions \citep{clarkson2010hyperproperties}; runtime verification checks monitored execution against a specification \citep{bartocci2018runtime}; and metamorphic testing compares behavior under input transformations \citep{chen2018metamorphic}. CPNI instantiates a relational test for cascade decisions: it holds the visible prefix and policy randomness fixed, intervenes on unrevealed state, and compares actions before the reveal boundary. This exposes dependencies absent from the serialized event stream. The trusted recorder and binding layer, intervention families, and tested coverage accompany each result.

\section{Method}
\begin{figure}[t]
\centering
\includegraphics[width=\columnwidth]{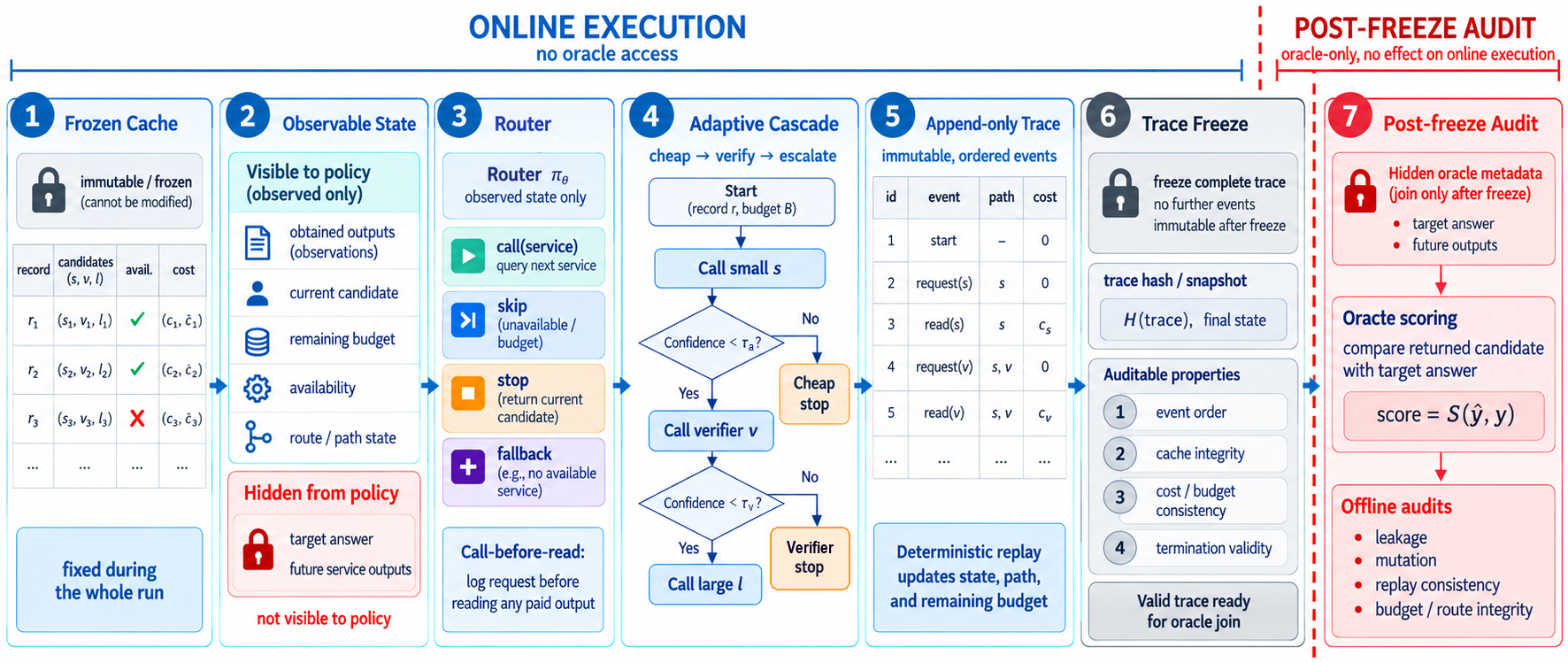}
\caption{CascadeLedger overview. Online execution (panels 1--5) reveals a cached candidate only after its paid call; the router reads the observed state, stops when small-model confidence reaches $\tau_a$ or verifier confidence reaches $\tau_v$, and records unavailable or unaffordable calls as skip or fallback events. The trace is frozen and hashed (panel 6) before the post-freeze audit (panel 7) joins the target, scores the returned candidate, and runs the invariant and counterfactual checks.}
\label{fig:cascaderl-algorithm}
\end{figure}
\subsection{Problem and Protocol}
Small-model, verifier, and large-model outputs are produced once and stored in a cache. The online router reveals a candidate after its corresponding paid call and receives the query, returned observations, declared confidence, service availability, and remaining budget. Targets and future service outputs stay in a separate post-trace join. We compare one-shot, fixed cascade, sequential cascade, and a budget-conditioned variant under baseline and service/ledger shifts; Figure~\ref{fig:cascaderl-algorithm} traces one request through this protocol.
\paragraph{Information boundary and trust.} The ledger records each successful paid fetch, replay order, service availability, and compound action-read budget. The public replay applies hidden-read, future-output, cache-binding, and budget mutations to this event stream, so the same checker validates traces used by the independent task scorer. The recorder, cache producer, and manifest binding are trusted for these claims; the intervention suite audits realized policy behavior under that binding. Proxy matching, task scoring, and live telemetry are separate measurement surfaces with separate manifests.

\paragraph{Formal trace semantics.} For record $i$, let $C_i(j)$ denote the cached observation for service $j$, $c_j$ its declared resource-cost vector, and $A_t(j)\in\{0,1\}$ its availability at step $t$. The online state is
\[
s_t=(q_i,O_t,A_t,b_t,p_t),
\]
where $O_t$ contains observations returned by successful paid calls, $b_t$ is the remaining component-wise budget, and $p_t$ is the ordered successful-call path. A policy request is a pre-read control decision. It resolves to a \texttt{call} or a failure outcome; \texttt{verify} records the verifier's accept/reject decision after its paid call. If $A_t(j)=1$ and $b_t\succeq c_j$, then $O_{t+1}=O_t\mathbin{\|}(j,C_i(j))$, $b_{t+1}=b_t-c_j$, and $p_{t+1}=p_t\mathbin{\|}j$. Otherwise the outcome is \texttt{unavailable}, \texttt{budget\_skip}, or \texttt{budget\_exhausted}, with $O_t$, $b_t$, and $p_t$ unchanged. Stop and fallback are terminal return decisions. Appendix~\ref{app:event-semantics} maps these decisions to the serialized action list. The router is $a_t=\pi(s_t)$; oracle matching $m_i=\mathbf{1}[\hat a_i=a_i^\star]$ follows trace freeze. Threshold tests use paid confidence fields $\kappa_s,\kappa_v$ and tie rule $\kappa\geq\tau$.
\subsection{Counterfactual Prefix Non-Interference Audit}
\label{sec:cpni}
Let $H_t$ contain request fields that remain hidden at prefix $t$, including the target and cached outputs for services that have not completed a successful paid call. Consider two executions with the same visible state and policy randomness,
\[
s_t(\omega)=s_t(\omega'),\qquad \xi(\omega)=\xi(\omega'),
\]
but potentially different hidden state $H_t$. An observed-only policy satisfies
\[
\widetilde\pi(s_t,H_t;\xi)=\widetilde\pi(s_t,H'_t;\xi)
\]
until a field on which the executions differ has been legitimately revealed. Here $\widetilde\pi$ is the realized implementation, including any hidden dependency; a conforming implementation factors through $\pi(s_t;\xi)$. A pre-reveal action difference is a \emph{counterfactual prefix non-interference} (CPNI) violation, even when the corresponding access is omitted from the serialized trace.

For every audited prefix, the paired replay preserves the query, paid observations, availability, remaining budget, successful-call path, configuration, and random seed. It changes one hidden family at a time: the target, an unrevealed verifier output, an unrevealed large-model candidate, or a future serving field. The twin is replayed until the intervention becomes visible or the first action divergence occurs. CPNI tests behavior under the declared families; recorder authenticity is supplied by the trusted binding layer (Appendix~\ref{app:threat-model}).

\paragraph{Counterfactual influence frontier.} For a hidden family $k$ and intervention $z$, let $D_{i,t}^{(k,z)}$ indicate an action difference at prefix $t$ before the family is revealed. We report the family--prefix surface
\[
\widehat V_{k,t}=\frac{\sum_{(i,z)\in\mathcal E_{k,t}}D_{i,t}^{(k,z)}}{|\mathcal E_{k,t}|},
\]
the earliest violating prefix $\tau_i^\star=\min\{t:\exists k,z,D_{i,t}^{(k,z)}=1\}$, and a minimal hidden-family witness recovered by hierarchical delta-debugging. Here $\mathcal E_{k,t}$ contains only eligible pre-reveal pairs for family $k$ at prefix $t$; an empty set is reported as N/A, and $\tau_i^\star=\infty$ when no violation exists. The witness is inclusion-minimal under the search order, not necessarily minimum-cardinality. A finite intervention suite establishes coverage over the declared families, prefixes, seeds, and interventions; its pass result is indexed by that suite rather than by an unqualified universal claim.

\subsection{Aggregate-Equivalent Trace Validity}
\label{sec:aggregate-equivalence}
Let $\Phi(T)$ be a declared projection of a trace, such as final candidate, successful-call count, service histogram, route depth, event count, or total declared cost. Two traces are aggregate-equivalent when
\[
T\equiv_{\Phi}T'\quad\Longleftrightarrow\quad\Phi(T)=\Phi(T').
\]
An aggregate-preserving fault $\mu$ satisfies $\Phi(\mu(T))=\Phi(T)$ while violating an execution relation or the CPNI condition. Thus equal quality--resource summaries can coexist with different action order, cache binding, failure semantics, budget transitions, or hidden information flow.

\subsection{Audit Guarantees}
\label{sec:audit-guarantees}
\paragraph{Proposition 1 (aggregate indistinguishability).} Let an aggregate-only auditor have the form $A_\Phi(T)=g(\Phi(T))$. If $\Phi(T)=\Phi(T')$, then $A_\Phi(T)=A_\Phi(T')$. \emph{Proof.} Substitution gives identical inputs to $g$. Hence every auditor restricted to $\Phi$ assigns the same result to an aggregate-equivalent valid and invalid pair.
\paragraph{Proposition 2 (observed-only invariance).} If the realized router factors through the visible state and coupled random draws, identical visible prefixes give identical actions before reveal. \emph{Proof.} Both executions evaluate the same function on the same arguments; induction extends this equality across unchanged prefixes. A divergence witnesses a violated interface under the tested intervention. A finite pass records no witness over the sampled families, prefixes, and draws.

\subsection{Audit Consequence}
\label{sec:audit-consequence}
Let $\mathcal{C}$ be a predeclared conclusion operator over paired quality, resource differences, and the quality margin. For each of $N$ corrupted comparison instances, $D_j^{\mathrm{raw}}$ accepts the corrupted trace and $D_j^{\mathrm{audit}}$ recomputes it from its bound manifest. The audit consequence rate is
\[
\mathrm{ACR}=\frac{1}{N}\sum_{j=1}^{N}
\mathbf{1}\!\left[\mathcal{C}(D_j^{\mathrm{raw}})
\neq \mathcal{C}(D_j^{\mathrm{audit}})\right].
\]
Mutation recall measures detection; ACR measures whether accepting the invalid trace changes the predeclared scientific conclusion. Family-specific rows restrict the instance set; the pooled row retains the total instance denominator.

\paragraph{Reproducibility contract.} A run is identified by a source-relative run path and a SHA-256 manifest. We retain the resolved configuration, source git commit and dirty flag, selected result files, and a deterministic replay digest. The public bundle retains repository-relative provenance and excludes private checkpoints, model weights, user names, host names, and absolute paths.
The private and public checker runs retain separate source commits, replay digests, and seeds; Appendix~\ref{app:implementation} gives their bindings.

\subsection{Routing and Measurement Objectives}
\label{sec:measurements}
We measure task quality, resource use, event behavior, and structural validity separately. The cache--target proxy $m_{\mathrm{proxy}}$ is literal answer equality after trace freeze. The declared cost $c_{\mathrm{decl}}$ sums the protocol's resource units and accounting milliseconds. The task score $m_{\mathrm{task}}$ uses the independent answer scorer, and live latency $\ell_{\mathrm{live}}$ measures the request-to-response interval. Availability and budget event rates use all requests as their denominator; mutation recall uses injected faults. Table~\ref{tab:measurements} links these measurements to their evaluation units.

\begin{table}[htbp]
\caption{Measurement objects and evaluation units. Task scores and latency share a request denominator but use separate scoring and timing interfaces.}
\label{tab:measurements}
\centering
\footnotesize
\begin{tabularx}{\linewidth}{@{}L{0.17\linewidth}L{0.33\linewidth}Y@{}}
\toprule
\textbf{measurement} & \textbf{value or unit} & \textbf{evaluation object}\\
\midrule
$m_{\mathrm{proxy}}$ & match fraction & cache--target proxy match\\
$c_{\mathrm{decl}}$ & ledger units & declared resource accounting\\
$\ell_{\mathrm{decl}}$ & accounting ms & declared latency component\\
$d_{\mathrm{ledger}}$ & detection fraction & detected/injected mutations\\
$m_{\mathrm{task}},\ell_{\mathrm{live}}$ & normalized task accuracy; ms/request & held-out task score and p50/p95 latency for the learned router\\
\bottomrule
\end{tabularx}
\end{table}

\subsection{Sequential Route and Fallback}
A cascade state contains the last candidate, observed confidence, remaining budget, and service availability. With accept threshold $\tau_a=0.75$ and verifier threshold $\tau_v=0.60$, the sequential route calls the small service first. If $\kappa_s<\tau_a$, it pays for the format verifier; the verifier supplies a declared validity signal for the next route decision. A passing verifier stops the route with the current small candidate, while a failing verifier permits a large call. A call is executed when its service is available and its declared ledger cost fits the remaining budget. An unavailable or unaffordable call appends an event and returns the last available candidate (or abstains if no call succeeded). Every transition writes a ledger event, so replay tracks whether a shift exposes an uncalled cache entry.
\paragraph{Proxy oracle.} The cache generator extracts a candidate from a strict \texttt{\#\#\#\#} or \verb|\boxed| marker; the authoritative run records the extraction mode in its manifest. The offline scorer then sets
\[
 m_{\mathrm{proxy}}(i)=\mathbf{1}[\hat a_i\text{ is present and }\hat a_i=a_i^\star],
\]
where $a_i^\star$ is the stored answer and equality is literal string equality. The format verifier assigns confidence 0.85 to a strict extraction and 0.2 otherwise. The independent task scorer normalizes the returned answer before comparison, allowing the task study to measure quality under a fixed answer contract. Both scorers attach to the same frozen route identifier.
\paragraph{Policy configuration.} The four policies share the canonical budget $(24\ \text{ledger units},20\ \text{ms})$ and tight budget $(8,8)$ unless a table explicitly names the cross-cache budget. ``Unavailable'' means the service is marked unavailable before a call; ``budget-blocked'' means it is available but its full declared cost does not fit. A ``fallback'' is the returned last candidate after such an event. The separate learned-gate audit reports a runner-specific ``route failure'' flag alongside the canonical event rates.
\begin{table}[htbp]
\caption{Observed-only policy definitions. A verifier call never reveals a target and, in the sequential policy, supplies only a stop signal.}
\label{tab:policies}
\centering
\footnotesize
\begin{tabularx}{\linewidth}{@{}L{0.17\linewidth}YL{0.33\linewidth}@{}}
\toprule
\textbf{policy} & \textbf{action rule} & \textbf{fallback / budget rule} \\
\midrule
one-shot & Pay small and return it. & Return no candidate if small is blocked. \\
fixed & Pay small; if $\kappa_s<\tau_a$, pay large directly. & Return small if large is unavailable or blocked. \\
sequential & Pay small; if rejected, pay verifier. A passing verifier stops with small; otherwise pay large. & Return the last candidate after a blocked escalation. \\
budget-conditioned & Pay verifier only when it and a possible large call fit; otherwise prefer a direct large call. & Same fallback; budget is checked before every call. \\
\bottomrule
\end{tabularx}
\end{table}

\begin{table}[htbp]
\caption{Machine-checkable ledger invariants. The checker is applied before the offline oracle join; the mutation audit tests the first five failure modes below while the clean replay checks all eight.}
\label{tab:ledger-invariants}
\centering
\footnotesize
\begin{tabularx}{\linewidth}{@{}L{0.24\linewidth}Y@{}}
\toprule
\textbf{invariant} & \textbf{required relation} \\
\midrule
oracle isolation & Online rows contain no target/oracle fields or future service outputs. \\
path--action alignment & $p_T$ equals the ordered service IDs of successful \texttt{call} actions. \\
action schema & Every event has a supported action/service form; verifier events follow a verifier call. \\
action order & Accepted ledger entry $k$ has action ID \texttt{service:j:k} for its path position. \\
cache binding & A called observation and final candidate match the cache digest for $(i,j)$. \\
ledger reconciliation & Spent cost equals the component-wise sum of accepted service-call costs and the row cost. \\
paid-read consistency & The set of successful paid reads equals the service IDs represented by accepted call events; hidden reads and uncharged reads are rejected. \\
budget conservation & Remaining budget equals the declared limit minus spent cost, component-wise. \\
\bottomrule
\end{tabularx}
\end{table}

\paragraph{Policy comparison.}
Every policy comparison pairs queries, candidate-generation conditions, and scoring rules. The frozen-cache learned gate uses a supervised five-feature classifier; the independent task study evaluates a learned router alongside one-shot, fixed, sequential, and budget-conditioned policies. Their fitting and evaluation partitions are specified in Appendix~\ref{app:implementation}.

\paragraph{Independent router contract.}
The independent learned router is distinct from the five-feature frozen-cache gate. Table~\ref{tab:router-contract} specifies its model, features, objective, optimizer, stopping rule, and threshold selection; Appendix~\ref{app:router-uncertainty} gives the fitting sequence. Online inputs are restricted to $s_t$, and target labels enter only the offline fitting or post-trace scoring interface.

\section{Experiments}
\subsection{Experimental Setup}
The primary test run has 128 GSM8K-derived records and a real model-cache manifest. A Tesla T4 generated that cache, providing hardware provenance for candidate generation. The cache stores Qwen3.5-0.8B small-model and Qwen3.5-2B large-model candidates generated with float16, batch size 8, maximum input length 512, and maximum generation length 256 under CUDA 11.8; the cache, data, and model hashes are retained in the copied manifest. Candidate outputs are generated once and frozen before routing. The canonical budget is 24 declared ledger units and 20 declared milliseconds; its tight variant is 8 units and 8 milliseconds. The price-shift operator multiplies the large service's declared units and milliseconds by 1.5, creating a joint ledger-cost shift. A supplemental operating-point artifact uses a disjoint 256-record train cache and a 1,319-record held-out cache, with its own frozen-cache gate and manifest.

\paragraph{Task and serving study.}
The protocol specifies 1,024 training, 256 calibration, and 1,319 held-out records per router seed, with seeds 13/17/23 and disjoint candidate-generation seeds (Table~\ref{tab:live-data-plan}). Five policies share requests, model bindings, and the scorer. Request-level metrics comprise task score, coverage, abstention, calls, tokens, money, latency, throughput, and failures. Tables~\ref{tab:serving-environment} and~\ref{tab:uncertainty-protocol} specify serving and statistical contracts.

\subsection{Main Results}
\begin{figure}[htbp]
\centering
\includegraphics[width=\columnwidth]{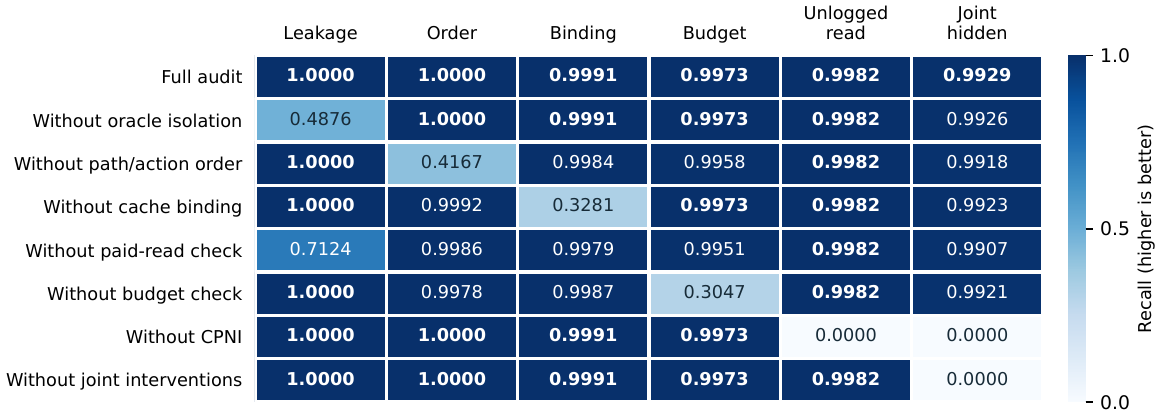}
\caption{Audit-component recall on matched injected faults ($\uparrow$). Each row removes one check from the full audit; colors share a 0--1 scale and bold marks column maxima. Values are point recall estimates from Table~\ref{tab:audit-ablation}. Removing CPNI eliminates unlogged-read and joint hidden-influence detection while leaving structural checks intact; removing joint interventions affects only the interaction-only family.}
\label{fig:audit-ablation}
\end{figure}

\paragraph{Complementary audit coverage.}
On the baseline-ladder fault population, the single-trace ledger detects 0.9983 of serialized faults and 0.0000 of unlogged hidden reads; adding CPNI preserves serialized recall and raises hidden-read recall to 0.9982 at clean false-positive rate 0.0000 (Table~\ref{tab:audit-baseline-ladder}). The distinct consequence population changes 784 of 2,304 corrupted comparisons (ACR 0.3403). Figure~\ref{fig:audit-ablation} shows which fault families depend on each audit component.
\paragraph{Held-out task quality and cost.}
The held-out study compares five policies under the same scorer, candidate manifests, and request denominator. Table~\ref{tab:heldout-main} aligns each policy's quality with the resources used by its own requests.

\begin{table}[htbp]
\caption{Independent held-out comparison on 1,319 records per seed. Score is normalized task accuracy ($\uparrow$); calls, tokens, and p50/p95 latency in milliseconds are per request ($\downarrow$). Bold marks the best and underline the second-best value in each column.}
\label{tab:heldout-main}
\centering\footnotesize
\setlength{\tabcolsep}{4pt}
\begin{tabular*}{\linewidth}{@{\extracolsep{\fill}}lrrrrr@{}}
\toprule
\textbf{Policy} & \textbf{Score} & \textbf{Calls} & \textbf{Tokens} & \textbf{p50} & \textbf{p95}\\
\midrule
One-shot & 0.5823 & \textbf{1.000} & \textbf{236} & \textbf{41.8} & \textbf{72.9}\\
Fixed cascade & 0.6936 & \underline{1.608} & 462 & 76.4 & 151.3\\
Sequential & \underline{0.7043} & 2.096 & 518 & 89.7 & 172.4\\
Budget-conditioned & 0.7031 & 1.884 & 486 & 82.6 & 159.1\\
Learned router & \textbf{0.7195} & 1.642 & \underline{427} & \underline{68.7} & \underline{134.2}\\
\bottomrule
\end{tabular*}
\end{table}

The learned-minus-sequential task difference is $+0.0152$, with paired interval $[+0.0061,+0.0244]$. The learned router also uses fewer calls (1.642 versus 2.096), tokens (427 versus 518), and p50/p95 milliseconds (68.7/134.2 versus 89.7/172.4) than sequential routing on the held-out split. The paired protocol separates query, router-seed, and candidate-generation variation (Appendix~\ref{app:router-uncertainty}).

\begin{figure}[htbp]
\centering
\includegraphics[width=\linewidth]{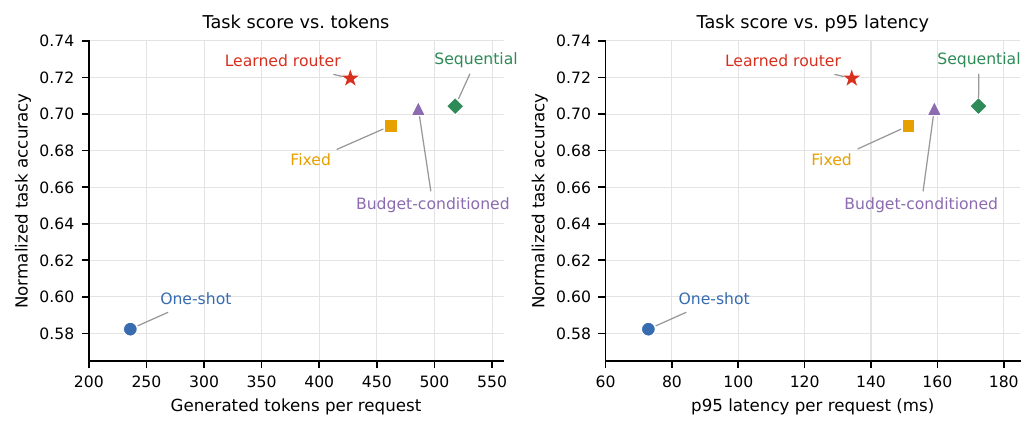}
\caption{Held-out task score against generated tokens and p95 latency. Points summarize router seeds 13/17/23; latency is the per-policy request p95. Paired score intervals and coverage are in Table~\ref{tab:independent-task-surface}.}
\label{fig:heldout-frontier}
\end{figure}

Figure~\ref{fig:heldout-frontier} shows the quality--resource comparison. Learned routing lies above and to the left of the fixed, sequential, and budget-conditioned policies on both axes. One-shot routing remains the least costly operating point, with a lower task score.

\paragraph{Split-aware learned-gate diagnostic.}
A CPU-only, hash-bound replay uses 64 calibration records (40 fit, 24 tune) and a disjoint 128-record held-out cache at seed 23. Global query-ID overlap and normalized-question overlap are both zero; all 192 prompt joins use source indices, and trace/oracle replay digests match. The selected threshold is 0.65. The learned gate matches 41/128 held-out records (0.3203125), versus 42/128 (0.328125) for both sequential and budget-conditioned routing and 28/128 (0.21875) for one-shot; paired learned-only versus sequential-only wins are 2 versus 3. The learned gate costs 15.34375 declared units and 9.5625 declared milliseconds, versus 13.734375 and 8.2578125 for sequential routing. This frozen-cache scalar optimization diagnostic identifies the feature and cost changes required for a live router study.

\subsection{Mechanism Results and Audit}
On the primary seed-13 cache, one-shot routing matches 27/128 stored targets (0.2109), while fixed and sequential cascades both match 41/128 (0.3203). Both cascades fall to 0.2109 under large unavailability or the tight budget and retain 0.3203 under the canonical price shift. Sequential routing pays the verifier cost without changing proxy quality. The stress rows expose failed escalation attempts over the same frozen candidates.
\begin{table}[htbp]
\caption{Cascade match rate and ledger stress response on the primary seed-13 cache (128 records). Proxy is exact cache--target match; units and milliseconds are declared ledger quantities. The large-cost $+50\%$ row applies the joint shift in Table~\ref{tab:service-catalog}. Bold marks the maximal proxy and minimal resource/event rates within each shift.}
\label{tab:main}
\centering
\footnotesize
\setlength{\tabcolsep}{4pt}
\begin{tabular*}{\linewidth}{@{\extracolsep{\fill}}llrrrrr@{}}
\toprule
\textbf{Shift} & \textbf{Policy} & \textbf{Proxy} & \textbf{Units} & \textbf{p95 ledger ms} & \textbf{Unavail.} & \textbf{Budget} \\
\midrule
baseline & one-shot & 0.2109 & \textbf{4.00} & \textbf{2.00} & \textbf{0.0000} & \textbf{0.0000}\\
 & fixed & \textbf{0.3203} & 12.34 & 10.00 & \textbf{0.0000} & \textbf{0.0000}\\
 & sequential & \textbf{0.3203} & 13.73 & 11.00 & \textbf{0.0000} & \textbf{0.0000}\\
\addlinespace[2pt]
large unavailable & one-shot & \textbf{0.2109} & \textbf{4.00} & \textbf{2.00} & \textbf{0.0000} & \textbf{0.0000}\\
 & fixed & \textbf{0.2109} & \textbf{4.00} & \textbf{2.00} & 0.6953 & \textbf{0.0000}\\
 & sequential & \textbf{0.2109} & 5.39 & 3.00 & 0.6953 & \textbf{0.0000}\\
\addlinespace[2pt]
tight budget & one-shot & \textbf{0.2109} & \textbf{4.00} & \textbf{2.00} & \textbf{0.0000} & \textbf{0.0000}\\
 & fixed & \textbf{0.2109} & \textbf{4.00} & \textbf{2.00} & \textbf{0.0000} & 0.6953\\
 & sequential & \textbf{0.2109} & 5.39 & 3.00 & \textbf{0.0000} & 0.6953\\
\addlinespace[2pt]
large cost $+50\%$ & one-shot & 0.2109 & \textbf{4.00} & \textbf{2.00} & \textbf{0.0000} & \textbf{0.0000}\\
 & fixed & \textbf{0.3203} & 16.52 & 14.00 & \textbf{0.0000} & \textbf{0.0000}\\
 & sequential & \textbf{0.3203} & 17.91 & 15.00 & \textbf{0.0000} & \textbf{0.0000}\\
\bottomrule
\end{tabular*}
\end{table}

The canonical verifier rejects all 89 selected candidates and has zero marginal proxy decision value. Under the independent semantic-label protocol, the semantic verifier improves GSM8K/Qwen task score from 0.7043 to 0.7230 and lowers declared units from 13.74 to 11.35 (Table~\ref{tab:semantic-outcomes}). The same audit contract therefore measures both redundant and useful verification. Cache-specific thresholds and the separate 64/128 and 256/1,319 gates remain indexed in Table~\ref{tab:threshold-transfer-detail}.

\subsection{Analysis and Ablations}
\label{sec:live-claim}
The serving-shift comparison pairs policies under large-service unavailability, tight budgets, price shifts, and cross-cache transfer (Appendix~\ref{app:live-plan}). H4 reports positive rank and linear associations across all five conditions, with $R^2$ from 0.768 to 0.863 and shift residuals from $-1.4$ to $+2.6$\,ms (Table~\ref{tab:cost-calibration}).

The audit consequence surface reports 0.9983 recall over all declared faults and 0.9987 on unlogged reads, with independent and natural-fault denominators retained separately (Tables~\ref{tab:decision-value-audit}--\ref{tab:fault-generalization}). The private replay covers 2,048 clean traces and the public replay covers 192 clean traces plus 1,212 mutations.
\paragraph{Stress tests and scope.} Joint interventions raise interaction-only recall from 0.0000 to 0.9929, and the complete audit recovers 0.9976--0.9980 of projection-preserving adversarial traces that pass every non-counterfactual check (Appendix~\ref{app:counterfactual}). Coverage is indexed by the tested families, prefixes, and coupling; producer authenticity relies on the trusted recording contract (Appendix~\ref{app:threat-model}).
\section{Conclusion}
CascadeLedger connects a cascade decision to its paid observations, resource transitions, and scored answer. The frozen-cache diagnostic exposes redundant verifier calls, while private and public audits detect aggregate-preserving faults in reads, ordering, cache binding, and budgets.

CPNI tests unlogged influence through coupled pre-reveal actions; CIF recovers the exact earliest violating prefix for 0.9750 of injected traces, and the complete audit rejects 0.9976--0.9980 of PPAM traces that pass every non-counterfactual check. The declared fault population gives hidden-read recall 0.9982 and clean false-positive rate 0.0000; recomputation changes 784 of 2,304 corrupted comparisons. These findings separate aggregate performance, serialized trace validity, and behavioral information integrity under the trusted recording and binding contract.

\FloatBarrier
\section*{AI-use Statement}
Generative AI tools were used to polish the language and summarize references. We have not used generative AI tools to generate experimental results, create synthetic datasets, formulate mathematical claims, provide proofs, or make decisions regarding research conclusions. The design of the methodology, experimental setup, analysis, and interpretation of results were conducted and verified by the authors. Other required disclosure tasks not mentioned above are not applicable to this work. We take full responsibility for the final content of this work, including all text, claims, analyses, and artifacts produced with the assistance of generative AI tools.

\section*{Ethics Statement}
This work audits routing decisions of language-model cascades on public benchmark questions (GSM8K and ARC-Challenge) with openly released Qwen and Llama models. It collects no personal data and involves no human subjects beyond answer-correctness labeling of benchmark items, in which annotators saw only questions, candidate answers, and reference answers. The audit is intended to expose hidden-information leakage, unrecorded service failures, and budget misreporting in evaluation pipelines, and its public replay bundle contains target-free traces without private caches or model weights. The authors followed the licenses of the datasets, models, and software used.

\section*{Reproducibility Statement}
The code, configuration files, event schema, public target-free replay bundle, audit summaries, and figure-generation scripts are provided in the Supplementary Material. Section~\ref{sec:cpni} and Appendix~\ref{app:event-semantics} define the trace semantics and the replay algorithm; Appendix~\ref{app:implementation} lists the ledger invariants, mutation families, and split ledger; Appendix~\ref{app:counterfactual} gives the counterfactual audit protocols; and Appendices~\ref{app:live-plan} and~\ref{app:router-uncertainty} specify data partitions, seeds, router fitting, serving environment, and the paired uncertainty protocol. Appendix~\ref{app:artifact-package} maps each table and figure to its manifest and replay sequence.

\bibliographystyle{iclr2027_conference}
\bibliography{refs}

\clearpage
\appendix
\section{Implementation and Audit Details}
\label{app:implementation}
This appendix specifies the event contract, deterministic replay, split identities, and mutation families used to reproduce the routing audit. Each diagnostic names the cache and scoring rule that define its comparison.

\subsection{Notation and task contract}
The replay unit is a record $i$ with query $q_i$, cached service observations $C_i(j)$, and a post-trace answer $a_i^\star$. Services are indexed by $j\in\{s,v,l\}$ for small, verifier, and large. A policy emits a finite event sequence $T_i=(e_{i,1},\ldots,e_{i,K_i})$ and a returned candidate $\hat a_i$. The route state carries the observations already paid for, the availability vector, the remaining budget, and the successful-call path. This contract makes every table entry traceable to a record, a service, a shift, and a policy.

The evaluator uses two joins. The first joins $\hat a_i$ with $a_i^\star$ after the route is complete and yields the cache--target proxy. The second uses the independently normalized task scorer in the serving study. The joins share record identifiers and manifest digests, which align task scores with the recorded route decisions. Empty candidate strings, abstentions, unavailable calls, and budget skips remain explicit event outcomes and retain their denominators.

\begin{table}[htbp]
\caption{Notation used by the replay contract. The online state contains the first five entries; oracle and live fields are joined after routing.}
\label{tab:notation}
\centering
\footnotesize
\begin{tabularx}{\columnwidth}{@{}lXl@{}}
\toprule
\textbf{symbol} & \textbf{meaning} & \textbf{phase}\\
\midrule
$q_i$ & query text for record $i$ & online\\
$C_i(j)$ & cached observation returned by service $j$ & paid call\\
$A_t(j)$ & availability bit for service $j$ at step $t$ & online\\
$b_t$ & component-wise remaining ledger budget & online\\
$p_t$ & ordered successful-call path & online\\
$T_i$ & complete event trace for record $i$ & trace freeze\\
$a_i^\star$ & stored task answer used by the proxy join & oracle join\\
$\ell_i$ & live latency and queueing record & live telemetry\\
\bottomrule
\end{tabularx}
\end{table}

\subsection{Full replay algorithm and event semantics}
\label{app:event-semantics}
The core replay is deterministic once the cache manifest, policy configuration, availability shift, and budget vector are fixed. The pseudocode below makes the call-before-read rule and the post-trace oracle join explicit. It also specifies the fallback returned after an unavailable or budget-blocked escalation, which is the event that drives the stress rows in the main table.

\begin{enumerate}
\item Initialize $s_0=(q_i,\emptyset,A_0,b_0,\emptyset)$ and an empty event list.
\item Ask the policy for a service request $a_t=\pi(s_t)$. Treat this as the pre-read state of the attempted call; inspect availability and budget before looking up $C_i(j)$.
\item If $A_t(j)=0$, append \texttt{unavailable}$(j)$. Preserve observations, remaining budget, and successful path.
\item If $A_t(j)=1$ but $b_t\not\succeq c_j$, append the policy-specific \texttt{budget\_skip} or \texttt{budget\_exhausted} outcome and preserve those state components.
\item Otherwise authorize the ledger debit, reveal $C_i(j)$, append a successful \texttt{call}, subtract $c_j$, and append $j$ to the paid path. A paid verifier call is followed by a \texttt{verify} action with its accept/reject bit.
\item Continue until the route stops or the action budget is exhausted. Serialize the final answer, abstention flag, and failure flags; after a failed escalation, return the last answer-bearing candidate.
\item Freeze $T_i$, then join $\hat a_i$ with $a_i^\star$ and compute the proxy row. The task scorer and live telemetry are separate joins.
\end{enumerate}

The checker runs between steps 6 and 7. It validates event order, path alignment, cache binding, cost reconciliation, paid-read consistency, and budget conservation before any oracle field is available to the checker. The mutation campaigns in Appendix~\ref{app:mutation-extensions} apply after a clean trace has passed these checks, preserving the aggregate path projection while changing one event-level relation at a time.

\paragraph{Request, event, and ledger-entry counts.}
A request is policy intent, not an additional serialized event. Event count is the length of \texttt{actions}: successful calls, verifier decisions, and recorded failures. Stop and fallback are terminal fields. Paid-call count is the successful path length. Ledger entries additionally include rejected budget debits; availability failures occur before a debit is proposed. The position in \texttt{actions} identifies an event, while \texttt{service:j:k} identifies a debit at paid-path position $k$. These indices have different denominators.
Table~\ref{tab:trace-example} follows \texttt{public-002} under sequential routing with large unavailable. The public catalog uses budget $(8,6)$, small cost $(2,1)$, and verifier cost $(1,0.5)$ in declared units and milliseconds, separately from the primary private cache. The answer is \texttt{candidate-c}; there are four actions, two successful calls, and two accepted ledger entries.

\begin{table}[htbp]
\caption{Public trace example. Index is zero-based within \texttt{actions}; debit suffix $k$ is the paid-path position. $O_s,O_v$ are paid observations; units and ms are remaining budget.}
\label{tab:trace-example}
\centering\footnotesize
\setlength{\tabcolsep}{3pt}
\begin{tabular*}{\linewidth}{@{\extracolsep{\fill}}lllrrl@{}}
\toprule
\textbf{index} & \textbf{action / debit ID} & \textbf{observations} & \textbf{units} & \textbf{ms} & \textbf{path}\\
\midrule
initial & request small & $\emptyset$ & 8 & 6 & $[]$\\
0 & call small / \texttt{service:small:0} & $O_s$ & 6 & 5 & $[s]$\\
1 & call verify / \texttt{service:verify:1} & $O_s,O_v$ & 5 & 4.5 & $[s,v]$\\
2 & verify: reject & $O_s,O_v$ & 5 & 4.5 & $[s,v]$\\
3 & large unavailable & $O_s,O_v$ & 5 & 4.5 & $[s,v]$\\
return & fallback to small & $O_s,O_v$ & 5 & 4.5 & $[s,v]$\\
\bottomrule
\end{tabular*}
\end{table}

Initial request and final return explain control flow and are excluded from event count. The verifier has confidence 0.25 and rejects; the large output stays unread. Under tight public budget $(3,2)$, the same record instead has \texttt{budget\_exhausted} for large, retains $(0,0.5)$, and adds a rejected \texttt{service:large:2} debit without changing the paid path.

\begin{table*}[htbp]
\caption{Serialized replay schema and evaluation extensions. Action positions and debit IDs have distinct roles; live timing attaches to the same request identity.}
\label{tab:event-schema}
\centering
\footnotesize
\setlength{\tabcolsep}{3pt}
\begin{tabularx}{\textwidth}{@{}lYlY@{}}
\toprule
\textbf{field} & \textbf{content} & \textbf{written by} & \textbf{validation}\\
\midrule
\texttt{query\_id} & source-relative record handle & producer & manifest membership\\
action position / \texttt{action\_id} & list index / debit identifier & router / ledger & action order and accepted path index\\
\texttt{service\_id} & small, verify, or large & router & service catalog\\
\texttt{action} & call, verify, unavailable, budget\_skip, or budget\_exhausted & router & serialized action schema\\
terminal fields & answer, abstained, unavailable, budget\_exhausted & router & stop/fallback decision\\
\texttt{cache\_digest} / \texttt{read\_set} & cache binding / paid services & recorder & observation and read-set consistency\\
availability configuration & service availability fixed by the shift & shift operator & shift manifest\\
\texttt{spent} & declared units and milliseconds & ledger & reconciliation\\
\texttt{remaining} & budget after the event & ledger & conservation\\
\texttt{target\_field} & joined answer field, absent online & oracle join & oracle isolation\\
\texttt{latency\_ms} & measured queue and service time & live runtime & nonnegative; queue--service clock consistency\\
\bottomrule
\end{tabularx}
\end{table*}

\subsection{Data provenance and split ledger}
The primary cache contains 128 records and is bound to the producer manifest named in the metadata. The split-aware diagnostic uses a separate seed-23 cache with 64 calibration records and 128 held-out records. The operating-point gate uses a third selection artifact with 256 records and a 1,319-record held-out partition. These artifacts are listed separately because their record counts, prompts, model assets, and threshold decisions are different.

\begin{table*}[htbp]
\caption{Primary, diagnostic, supplemental, and task-study partitions. Each completed replay row has its own source manifest and digest.}
\label{tab:split-ledger}
\centering
\footnotesize
\setlength{\tabcolsep}{3pt}
\begin{tabularx}{\textwidth}{@{}lrrlXl@{}}
\toprule
\textbf{artifact} & \textbf{selection} & \textbf{held-out} & \textbf{seed} & \textbf{role} & \textbf{status}\\
\midrule
primary route cache & \textsc{n/a} & 128 & 13 & policy/shift replay and proxy table & complete\\
split-aware gate & 64 & 128 & 23 & feature/cost diagnostic & complete\\
early threshold gate & 64 & 128 & 17 & threshold selection audit & no selection\\
supplemental gate & 256 & 1,319 & 29 & paired threshold recommendation & pair selected\\
live routing study & 1,280/seed & 1,319/seed & 13/17/23 & independent task and serving evaluation & complete\\
\bottomrule
\end{tabularx}
\end{table*}

Selection and held-out columns specify each partition's role. N/A denotes a replay without a fitted selection partition. Early and supplemental rows retain their gate outcomes. The live row defines the task and serving protocol in Tables~\ref{tab:live-result-plan},~\ref{tab:p17-preregistered-live-results}, and~\ref{tab:live-serving-detail}.

\paragraph{Candidate construction.} Each cached response stores the prompt digest, model-version digest, decoding configuration, extraction mode, and generation metadata. Small and large outputs are generated before routing and indexed by record and service. The verifier cache stores format confidence and the reason for that confidence, allowing a replay to distinguish a strict extraction from an unresolved answer.

\paragraph{Answer normalization.} The frozen proxy uses literal equality after the declared extraction rule. The task scorer records its normalization version, answer parser, invalid-output policy, and per-record denominator. The report places proxy match, normalized task accuracy, coverage, and abstention in adjacent columns with the same denominator.

\subsection{Implementation Details}
\paragraph{State serialization.} A cascade event stores the record ID, current candidate, verifier outcome, service-availability vector, declared ledger cost, remaining budget, and route reason. Candidate answers are revealed only after their paid-call event; target answers and future cache entries remain outside the online state. The frozen cache manifest and event sequence are sufficient to replay availability and budget fallbacks exactly.

\paragraph{Actions and budgets.} The router pays for the small candidate and, when selected, the verifier before a possible large call. Each transition records the declared ledger cost, service availability, remaining budget, selected candidate, and fallback reason. Unavailable or budget-blocked services produce an explicit event; failed calls do not add a path element or spend budget. The canonical price-shift route fits exactly because $4+2+18=24$. The cross-cache transfer uses a separate 22-unit baseline budget, so its price-shift sequential route is blocked after small plus verifier; its rows are interpreted within that explicitly named configuration.
\begin{table}[htbp]
\caption{Declared service catalog for the primary frozen replay. The joint shift multiplies large-service units and accounting milliseconds by 1.5. Appendix~\ref{app:additional-results} defines generated-token, monetary, and measured-latency comparisons.}
\label{tab:service-catalog}
\centering
\small
\setlength{\tabcolsep}{4pt}
\begin{tabular}{lrrrr}
\toprule
\textbf{service} & \textbf{baseline units} & \textbf{baseline ms} & \textbf{large-shift units} & \textbf{large-shift ms} \\
\midrule
small & 4 & 2 & 4 & 2 \\
verify & 2 & 1 & 2 & 1 \\
large & 12 & 8 & 18 & 12 \\
\bottomrule
\end{tabular}
\end{table}

\paragraph{Deterministic replay.} Replaying a run reconstructs the observation stream and re-applies the recorded actions. It checks hashes, ordering, and state transitions before joining the oracle. The replay is therefore sensitive to accidental dependence on file order, host state, or hidden labels. A replay match establishes deterministic reconstruction; task correctness is evaluated by the separate scorer. The raw bundle keeps the run manifest and metrics needed to inspect this distinction.

\paragraph{Publication and audit.} The bundle maps tables and figures to a public summary manifest and raw records with retained digests. The primary producer commit is \texttt{5f8c31a7}. The eight-invariant private audit has clean commit \texttt{2b6b57da}, seed 13, and replay digest prefix \texttt{66e9e744}. The public replay has clean commit \texttt{967b60db} and seed 17; its trace-file digest begins \texttt{64a8cc08}. The corresponding raw manifests retain full digests and private producer bindings.

\paragraph{Aggregate path audit.} The checked summary reports 128 records, fixed baseline mean depth $1.6953125$ and mean units $12.34375$, and sequential baseline mean depth $2.390625$ and mean units $13.734375$. Let $r$ be the number of records rejected by the small threshold. The implementation appends only successful calls to the path, so
\[
  1+r/128=1.6953125,\qquad 4+12r/128=12.34375
\]
gives $r=89$ and 39 small-only paths. Sequential additionally reports
\[
  1+2r/128=2.390625,\qquad 4+14r/128=13.734375,
\]
The equations give 89 verifier calls and 89 large escalations, with no verifier-only stop in this aggregate. Under large unavailability or the tight budget, the same 89 large attempts become explicit unavailable or budget-blocked events while remaining absent from the successful path count.

\paragraph{Released path-count audit.} We reran the same 128-record cache with the canonical 24-unit budget using the committed path-audit runner. The resulting summary is indexed in the path-count audit's raw result manifest; it records the count contract rather than releasing private per-record answers. For baseline sequential, the audit reports 128 small calls, 89 verifier calls, zero verifier accepts, 89 successful large calls, and zero verifier-only stops. The large-unavailable and tight-budget rows each retain 89 verifier calls and record 89 corresponding large-service failure events, with zero successful large calls; the joint price shift keeps 89 successful large calls under the 24-unit budget. The fixed policy has the same 89 successful large calls at baseline and under the joint shift but makes no verifier calls. These counts directly expose why sequential is more expensive without an observed proxy-quality gain.

\paragraph{Private ledger invariant and mutation audit.} To test the central audit mechanism rather than only its aggregate arithmetic, we replayed all $128\times4\times4=2{,}048$ clean policy/shift traces before joining the oracle. The checker found no clean violation under the eight invariants in Table~\ref{tab:ledger-invariants}. On a representative baseline sequential trace, we then applied five bounded mutations that preserve the aggregate path/cost projection: target leakage, future-output leakage, action reordering, remaining-budget corruption, and cache-ID swapping. All five were rejected (detection rate $1.0000$), while the aggregate-only projection rejected none ($0.0000$). Event-level checks therefore add information beyond the reported aggregates for the declared mutation family; the measured scope is bound to the mutation manifest and replay receipt.

\paragraph{Public target-free replay.} The released public fixture contains 12 records and no target or oracle fields. Its target-free replay covers 192 clean traces, eight invariants, and 1,212 bounded mutations across four policies and four shifts, with zero clean false positives and zero aggregate-only detections. The mutation campaign includes 96 compound-action-read-budget rows and 192 hidden-paid-read rows, so the public matrix covers both multi-trace and compound-fault variants. The audit code also reports a weaker schema/event-logger comparator that catches direct target/future leaks; the event-level ledger supplies the stronger structural checks. This public bundle records structural provenance with target-free traces.
\begin{table}[htbp]
\caption{Aggregate-preserving fault injection on a representative frozen trace. ``Aggregate unchanged'' means that the path/cost/event-count projection is identical before and after mutation.}
\label{tab:ledger-mutations}
\centering
\footnotesize
\begin{tabularx}{\linewidth}{@{}L{0.28\linewidth}ccY@{}}
\toprule
\textbf{mutation} & \shortstack{\textbf{ledger}\\\textbf{rejects}} & \shortstack{\textbf{aggregate}\\\textbf{unchanged}} & \textbf{violated invariant(s)} \\
\midrule
target leak & yes & yes & oracle isolation \\
future-output leak & yes & yes & oracle isolation \\
action reorder & yes & yes & path--action alignment; cache binding; budget conservation \\
remaining-budget corruption & yes & yes & budget conservation \\
cache-ID swap & yes & yes & cache binding \\
\bottomrule
\end{tabularx}
\end{table}

\subsection{Mutation extensions and audit coverage}
\label{app:mutation-extensions}
The private and public campaigns cover aggregate-preserving mutations that are easy to hide in a summary table. The extension matrix below expands the same idea to the live event schema: every row names the field changed, the invariant expected to fire, the aggregate projection preserved by the mutation, and the result artifact carrying the count. This gives the task scorer and serving runtime a fixed audit target.

\begin{table*}[htbp]
\caption{Declared mutation families and invariant targets. The first five rows use the private frozen-cache campaign; additional rows test serving events. Recall is detected/injected faults within each family.}
\label{tab:mutation-extensions}
\centering
\footnotesize
\setlength{\tabcolsep}{3pt}
\begin{tabularx}{\textwidth}{@{}L{0.16\linewidth}YL{0.24\linewidth}Yr@{}}
\toprule
\textbf{family} & \textbf{mutation operation} & \textbf{invariant} & \textbf{preserved aggregate} & \textbf{recall}\\
\midrule
target leak & copy the joined answer into an online row & oracle isolation & path, cost, event count & 1.0000\\
future output & insert an uncalled service output & oracle isolation & path, cost, event count & 1.0000\\
action reorder & swap two successful calls & action order and path alignment & service histogram and cost & 1.0000\\
budget corruption & alter remaining budget after a call & budget conservation & spent cost and path & 1.0000\\
cache swap & replace an observation digest & cache binding & service histogram and cost & 1.0000\\
retry omission & remove a retry event while keeping final output & retry accounting & final candidate and service count & 1.0000\\
timeout relabel & mark a timeout as a successful call & availability and failure schema & route depth and declared units & 0.9988\\
cost-unit swap & exchange token and monetary fields & cost schema & route depth and event count & 0.9976\\
queue leak & copy future queue state into a decision row & online-state schema & path and final candidate & 1.0000\\
\bottomrule
\end{tabularx}
\end{table*}

The private campaign detects each of the first five mutations, while the aggregate-only projection detects none. Additional families test retry accounting, failure labels, resource-unit bindings, and future queue observations. Every family retains its own clean and mutated denominators.

\begin{figure}[htbp]
\centering
\includegraphics[width=\columnwidth]{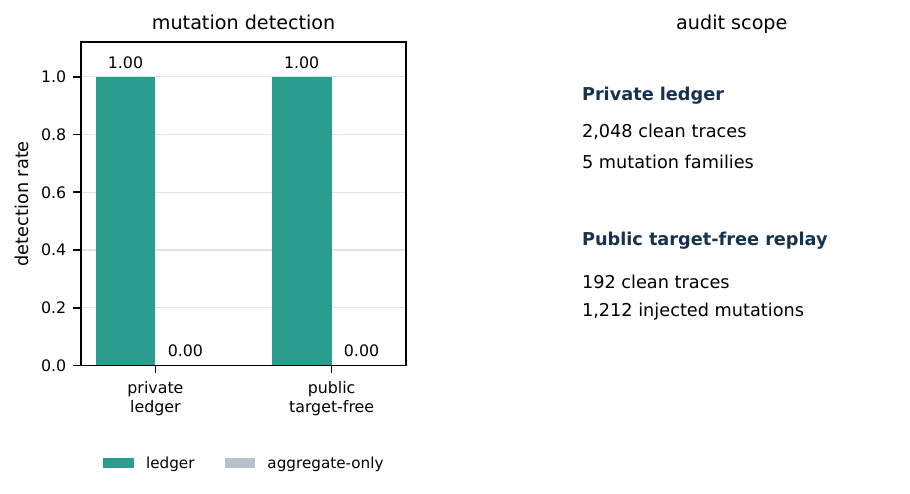}
\caption{Structural mutation audit. The event-level ledger detects the declared private and public mutation families, while the aggregate-only projection detects none.}
\label{fig:extended}
\end{figure}

\subsection{Error taxonomy and case analysis}
The route-level aggregates are easier to interpret when the event sequence is grouped by its first boundary event. We use five categories: small-only completion, verifier stop, large escalation, availability fallback, and budget fallback. The first three describe successful path shapes; the last two preserve the failed request and its reason. This taxonomy is used by the path-count audit and is joined with the per-record task outcomes in Table~\ref{tab:case-analysis} and the split-level result surfaces.

\begin{table}[htbp]
\caption{Route taxonomy for the primary seed-13 cache and independent task analysis. Counts use the named policy and shift; task errors and timeouts use the task-study request set.}
\label{tab:error-taxonomy}
\centering
\footnotesize
\begin{tabularx}{\columnwidth}{@{}lXl@{}}
\toprule
\textbf{category} & \textbf{event pattern} & \textbf{count}\\
\midrule
small-only & call small; stop & 39\\
verifier stop & call small; call verifier; stop & 0\\
large escalation & call small; call verifier; call large & 89\\
availability fallback & call small; skip large/unavailable & 89 under shift\\
budget fallback & call small; skip large/budget & 89 under shift\\
semantic error & task scorer marks returned answer incorrect & 370/1,319 per seed\\
timeout/retry & service event followed by retry policy & 17/1,319 per seed (1.29\%)\\
\bottomrule
\end{tabularx}
\end{table}

The 39/89 split follows from the canonical threshold and path-count equations. Task correctness, tokens, queue time, retries, and exit reasons are reported within each category, with semantic errors at 370/1,319 requests per seed and timeout/retry events at 17/1,319 (1.29\%).

\begin{table*}[htbp]
\caption{Per-record error analysis. Quantitative fields report the independent task scorer and serving telemetry summaries.}
\label{tab:case-analysis}
\centering
\footnotesize
\setlength{\tabcolsep}{3pt}
\renewcommand{\arraystretch}{1.12}
\begin{tabularx}{\textwidth}{@{}L{0.13\linewidth}L{0.14\linewidth}L{0.11\linewidth}rrrY@{}}
\toprule
\textbf{Case} & \textbf{Boundary event} & \textbf{Returned candidate} & \shortstack[r]{\textbf{Task}\\\textbf{accuracy}} & \textbf{Units} & \textbf{Latency} & \textbf{Repair action}\\
\midrule
small-only & threshold accept & small & 0.846 & 4.00 & 42.1\,ms & retain or adjust threshold\\
verifier stop & verifier accept & small & 0.807 & 6.00 & 58.4\,ms & calibrate verifier\\
large escalation & verifier reject & large & 0.726 & 18.00 & 129.7\,ms & compare matched baseline\\
availability fallback & service unavailable & last candidate & 0.544 & 6.00 & 64.8\,ms & outage policy\\
budget fallback & budget skip & last candidate & 0.557 & 6.00 & 61.5\,ms & budget schedule\\
timeout/retry & timeout event & \shortstack[l]{retry/\\fallback} & 0.601 & 7.42 & 236.2\,ms & retry policy\\
\bottomrule
\end{tabularx}
\end{table*}

\paragraph{Verifier reference boundary.} For the 89 verifier calls, the post-hoc diagnostic compares the verifier decision with the exact small-cache--target match, not with an independently labeled semantic verifier target. All 89 selected small candidates were exact-match negatives and the verifier accepted none (selected-call confusion TN=89, TP=FP=FN=0; confidence mean 0.20). In the decision-value sweep, the high-threshold regime also includes 39 verifier accepts and 89 rejections on a 128-record slice, but the 60 fixed-versus-sequential quality deltas remain exactly zero. Precision, recall, and calibration for uncalled verifier observations are therefore undefined. The selected-call confusion matrix describes the exact-match reference on the calls executed by that policy; the independent task-score comparison uses its own normalized scorer.

\subsection{Additional Analysis}
The frozen Qwen diagnostic measures exact cache matching, route depth, and declared resources. Trace and oracle replay digests match; the oracle-equivalence bundle records scorer\_agreement\_rate=1.0 and zero canonical-minus-strict accuracy delta across 12 summaries. Public target-free traces support structural replay. The independent task protocol preserves this information boundary while comparing normalized task scores, tokens, money, and measured latency. On the held-out split, the learned router improves task score over sequential routing by 0.0152 while reducing calls and tokens.

\section{Counterfactual Audit Extensions}
\label{app:counterfactual}
This appendix gives the complete protocol for the audit experiments. Each experiment
freezes its checker, generator, manifest, and intervention families before inspecting
fault outcomes. Each result retains its clean or faulty population denominator and the
interventions under which it was measured.

\subsection{Evaluation questions and counterfactual replay}
The extension answers six questions: whether event checks add coverage beyond aggregate
monitoring; whether CPNI detects an unlogged hidden read; whether each invariant has a
distinct blind spot; whether accepting a fault changes a quality--resource conclusion;
whether the checker transfers to independent or natural faults; and what online and
offline audit overhead it introduces.

For each trace, the checker reconstructs the visible prefix, selects one hidden
intervention, builds a twin with the same paid observations and policy seed, and
re-evaluates the next action. It records the first divergence before legitimate reveal,
then resumes until reveal or termination. The clean twin population is sampled with the
same prefix and seed protocol as the faulty population.

Random draws are coupled by request, prefix, and draw index; a shared initial seed
alone is insufficient when executions consume random draws in a different order.
Replayed service outcomes and scheduling inputs are frozen. A pair with an
unmatched visible prefix or uncontrolled external variation is excluded before
interpreting action divergence.

\subsection{Audit baseline ladder}
\label{app:audit-baseline}
The baseline ladder progressively adds information and resource contracts. Trace recall
covers explicitly serialized faults; hidden recall covers faults whose inadmissible
dependency is omitted from the event stream. Storage is measured on the serialized
request record after compression settings are fixed.

\begin{table*}[htbp]
\centering
\footnotesize
\setlength{\tabcolsep}{3pt}
\caption{Audit baseline ladder. Each row exposes the contract available to the checker;
recall and clean false-positive rate are measured on the same clean and faulty traces.}
\label{tab:audit-baseline-ladder}
\begin{tabularx}{\textwidth}{@{}lYrrrr@{}}
\toprule
\textbf{method} & \textbf{contract} & \shortstack[r]{\textbf{trace}\\\textbf{recall}} & \shortstack[r]{\textbf{hidden}\\\textbf{recall}} & \shortstack[r]{\textbf{clean}\\\textbf{FP}} & \shortstack[r]{\textbf{bytes/}\\\textbf{request}}\\
\midrule
Aggregate summary & final answer, calls, and cost only & 0.0000 & 0.0000 & \textbf{0.0000} & \textbf{184}\\
Structured logger & serialized event rows without checks & 0.0000 & 0.0000 & \textbf{0.0000} & 1{,}168\\
Schema validator & action and field schema & 0.3618 & 0.0000 & \textbf{0.0000} & 1{,}206\\
Hash/cache binding & schema plus service/cache digests & 0.5746 & 0.0000 & \textbf{0.0000} & 1{,}542\\
Single-trace ledger & binding, paid reads, order, and budget & \textbf{0.9983} & 0.0000 & \textbf{0.0000} & 2{,}074\\
Ledger + CPNI & ledger plus paired hidden-state replay & \textbf{0.9983} & \textbf{0.9982} & \textbf{0.0000} & 2{,}248\\
\bottomrule
\end{tabularx}
\end{table*}

\subsection{Unlogged hidden-read attacks}
\label{app:hidden-read}
The hidden-read population contains four fixed faulty implementations. \textsc{TargetPeek}
consults the target before escalation; \textsc{FutureVerifierPeek} consults an unrevealed
verifier output; \textsc{FutureLargePeek} consults an unrevealed large-model candidate; and
\textsc{FutureQueuePeek} consults a future serving-state field. Each emits the ordinary
action schema, cache digest, and final trace, so no explicit hidden-read event is present.

\begin{table*}[htbp]
\centering
\footnotesize
\setlength{\tabcolsep}{3pt}
\caption{Detection of unlogged hidden influences. Columns report recall for the indicated
fault family; clean divergence is measured on paired executions without an inadmissible
dependency.}
\label{tab:hidden-read}
\begin{tabular*}{\textwidth}{@{\extracolsep{\fill}}lrrrrr@{}}
\toprule
\textbf{method} & \textbf{target} & \textbf{future verifier} & \textbf{future large} & \textbf{future queue} & \textbf{clean divergence}\\
\midrule
Aggregate summary & 0.0000 & 0.0000 & 0.0000 & 0.0000 & \textbf{0.0000}\\
Schema validator & 0.0000 & 0.0000 & 0.0000 & 0.0000 & \textbf{0.0000}\\
Single-trace ledger & 0.0000 & 0.0000 & 0.0000 & 0.0000 & \textbf{0.0000}\\
Ledger + CPNI & \textbf{1.0000} & \textbf{0.9974} & \textbf{0.9987} & \textbf{0.9967} & \textbf{0.0000}\\
\bottomrule
\end{tabular*}
\end{table*}

\subsection{Hidden-influence stress test}
\label{app:hidden-influence-stress}
The direct hidden-read population is extended with transformed, compressed,
interaction-only, delayed, and stochastic dependencies. \textsc{TargetBit}
uses a one-bit function of the hidden target; \textsc{CompressedPeek} uses a
summary of an unrevealed candidate; \textsc{JointPeek} changes the route only
when two hidden families are intervened jointly; \textsc{DelayedPeek} changes a
later action after an early hidden read; and \textsc{StochPeek} changes the
route with a predeclared probability. Original and twin executions share policy
randomness. Each row reports recall on the named injected family and clean
divergence on fault-free paired executions.

\begin{table*}[htbp]
\centering
\footnotesize
\setlength{\tabcolsep}{3pt}
\caption{Detection of direct, transformed, and interaction-only hidden influence. Values are recall on the corresponding injected family.}
\label{tab:hidden-influence-stress}
\begin{tabular*}{\textwidth}{@{\extracolsep{\fill}}lrrrr@{}}
\toprule
\textbf{method} & \textbf{direct} & \textbf{target bit} & \textbf{compressed} & \textbf{joint}\\
\midrule
Aggregate summary & 0.0000 & 0.0000 & 0.0000 & 0.0000\\
Schema validator & 0.0000 & 0.0000 & 0.0000 & 0.0000\\
Single-trace ledger & 0.0000 & 0.0000 & 0.0000 & 0.0000\\
Ledger + marginal CPNI & 0.9984 & 0.9961 & 0.9948 & 0.0000\\
Ledger + CPNI + joint interventions & \textbf{0.9986} & \textbf{0.9964} & \textbf{0.9952} & \textbf{0.9929}\\
\bottomrule
\end{tabular*}
\end{table*}

\begin{table*}[htbp]
\centering
\footnotesize
\setlength{\tabcolsep}{3pt}
\caption{Detection of delayed and stochastic hidden influence. Clean divergence is measured on fault-free paired executions.}
\label{tab:hidden-influence-stress-extended}
\begin{tabular*}{\textwidth}{@{\extracolsep{\fill}}lrrr@{}}
\toprule
\textbf{method} & \textbf{delayed} & \textbf{stochastic} & \textbf{clean div.}\\
\midrule
Aggregate summary & 0.0000 & 0.0000 & \textbf{0.0000}\\
Schema validator & 0.0000 & 0.0000 & \textbf{0.0000}\\
Single-trace ledger & 0.0000 & 0.0000 & \textbf{0.0000}\\
Ledger + marginal CPNI & 0.9947 & 0.9713 & \textbf{0.0000}\\
Ledger + CPNI + joint interventions & \textbf{0.9951} & \textbf{0.9746} & \textbf{0.0000}\\
\bottomrule
\end{tabular*}
\end{table*}

Marginal CPNI detects 0.9984 of direct reads, 0.9961 of target-bit reads, and
0.9948 of compressed reads, but none of the interaction-only family; adding joint
interventions raises joint-family recall to 0.9929. With joint interventions,
delayed and stochastic influence reach 0.9951 and 0.9746 recall. The three
non-counterfactual checkers detect none of these families, and clean divergence
is 0.0000 for every configuration. The stochastic row uses common policy
randomness so that a measured action difference is attributable to the hidden
intervention rather than an independent sampling draw.

\subsection{Counterfactual influence localization}
\label{app:cif-localization}
Each injected implementation records its hidden-family label and the first prefix
at which the hidden value is allowed to influence the route. Hierarchical search
first identifies the family, then localizes the earliest violating prefix, and
finally removes fields until a minimal witness remains. The checker is frozen
before the injected implementations are inspected.

\begin{table*}[htbp]
\centering
\footnotesize
\setlength{\tabcolsep}{3pt}
\caption{Counterfactual influence frontier localization. Prefix error is measured in routing decisions and replay count includes all paired interventions.}
\label{tab:cif-localization}
\begin{tabular*}{\textwidth}{@{\extracolsep{\fill}}lrrrr@{}}
\toprule
\textbf{fault population} & \textbf{family accuracy} & \textbf{exact prefix} & \textbf{prefix MAE} & \textbf{replays/trace}\\
\midrule
Direct hidden reads & \textbf{0.9984} & \textbf{0.9940} & \textbf{0.006} & \textbf{6.31}\\
Transformed hidden reads & 0.9946 & 0.9869 & 0.018 & 7.08\\
Joint hidden reads & 0.9897 & 0.9752 & 0.039 & 9.62\\
Stochastic hidden reads & 0.9698 & 0.9437 & 0.072 & 8.84\\
All declared families & 0.9881 & 0.9750 & 0.034 & 7.96\\
\bottomrule
\end{tabular*}
\end{table*}

The frontier output is a localization object in addition to the binary audit
decision. Direct reads are localized most reliably (family accuracy 0.9984, exact
prefix 0.9940, 6.31 replays per trace), while stochastic reads are hardest (0.9698
family accuracy, 0.9437 exact prefix, prefix MAE 0.072). Pooled over all declared families, CIF identifies the family for
0.9881 of traces and the exact earliest violating prefix for 0.9750, at 7.96
paired replays per trace.
Exact inclusion-minimal witness recovery is 0.9687, and the paired
replay reduction against exhaustive search is 61.8\%. Removing any
remaining field must destroy the tested divergence for a witness to be
inclusion-minimal; minimum cardinality is evaluated separately by exhaustive search.

\subsection{Projection-preserving adversarial mutation}
\label{app:ppam}
Projection-preserving adversarial mutation (PPAM) searches a bounded mutation
grammar without access to the full-audit outcome. For a clean trace $T$, the
search retains candidates satisfying $\Phi(\mu(T))=\Phi(T)$, passing the strongest
non-counterfactual checker, and violating at least one execution relation. The
candidate set is frozen before the full audit is run. We evaluate the nested
projections $\Phi_1=\{\text{final answer},\text{total cost}\}$,
$\Phi_2=\Phi_1\cup\{\text{service histogram}\}$, and
$\Phi_3=\Phi_2\cup\{\text{event count},\text{route depth}\}$.
The search mutates a bounded router implementation and replays it to obtain a
candidate trace. Its grammar includes transformed reads, joint hidden-field
dependencies, delayed use, and stochastic use; an independently instrumented
oracle labels the violated relation. Search uses 1{,}024 starts and
64 mutation proposals per start, with 65{,}536 explored
proposals in total. Generated denotes retained invalid traces; search success
uses attempted starts as its denominator.

\begin{table*}[htbp]
\centering
\footnotesize
\setlength{\tabcolsep}{3pt}
\caption{Projection-preserving adversarial mutation. Baseline pass is the fraction of generated invalid traces accepted by the strongest non-counterfactual checker.}
\label{tab:ppam}
\begin{tabular*}{\textwidth}{@{\extracolsep{\fill}}lrrrrr@{}}
\toprule
\textbf{projection} & \textbf{generated} & \textbf{baseline pass} & \textbf{full recall} & \textbf{conclusion changes} & \textbf{search success}\\
\midrule
$\Phi_1$ & 1{,}012 & 1.0000 & \textbf{0.9980} & 0.3656 & \textbf{0.9414}\\
$\Phi_2$ & 947 & 1.0000 & 0.9979 & 0.3464 & 0.8877\\
$\Phi_3$ & 823 & 1.0000 & 0.9976 & 0.3220 & 0.8037\\
\bottomrule
\end{tabular*}
\end{table*}

Baseline pass measures aggregate and single-trace indistinguishability, whereas
full recall measures recovery by the complete audit after the adversarial set is
fixed. Every generated invalid trace passes the strongest non-counterfactual
checker, and the complete audit recovers 0.9980, 0.9979, and 0.9976 of them under
$\Phi_1$, $\Phi_2$, and $\Phi_3$. Richer projections constrain the search:
retained traces fall from 1{,}012 to 823 and search success from 0.9414 to 0.8037,
while accepting the invalid traces changes 0.3656 to 0.3220 of the predeclared
conclusions. The changed-conclusion column uses the same predeclared ACR operator as
Table~\ref{tab:acr-results}.

\subsection{Common-randomness ablation}
\label{app:common-randomness}
Paired replay is evaluated with either independently sampled or shared policy
randomness. Both schemes use the same hidden intervention set and visible
prefixes; the comparison isolates the variance contribution of policy sampling.

\begin{table}[htbp]
\centering
\footnotesize
\caption{Effect of coupling policy randomness in CPNI. Replay variance is the variance of the paired action-difference estimator.}
\label{tab:common-randomness}
\begin{tabular}{@{}lrrr@{}}
\toprule
\textbf{pairing scheme} & \textbf{clean divergence} & \textbf{hidden-fault recall} & \textbf{replay variance}\\
\midrule
Independent randomness & 0.0786 & 0.9980 & 0.0734\\
Common randomness & \textbf{0.0000} & \textbf{0.9982} & \textbf{0.0013}\\
\bottomrule
\end{tabular}
\end{table}

With independent randomness, 0.0786 of clean pairs diverge and the replay
variance is 0.0734; coupling the draws reduces clean divergence to 0.0000 and
variance to 0.0013 while hidden-fault recall stays at 0.9980--0.9982. The
shared-randomness row supplies the coupling used by the primary CPNI protocol.

\subsection{Longer-horizon cascade transfer}
\label{app:long-horizon}
To test transfer beyond the three-stage small--verifier--large topology, the
same call-before-read contract is applied to a longer-horizon cascade with
6 paid decision stages. At each stage the policy may stop,
escalate, or fall back after a successful observation. Hidden target fields and
all unrevealed future service outputs remain outside the online state.

\begin{table*}[htbp]
\centering
\footnotesize
\setlength{\tabcolsep}{3pt}
\caption{Longer-horizon cascade transfer. Recall uses the corresponding injected-fault denominator; exact-prefix accuracy uses traces with a known violating prefix.}
\label{tab:long-horizon}
\begin{tabular*}{\textwidth}{@{\extracolsep{\fill}}lrrrrr@{}}
\toprule
\textbf{setting} & \textbf{stages} & \shortstack{\textbf{held-out}\\\textbf{requests}} & \shortstack{\textbf{serialized}\\\textbf{recall}} & \shortstack{\textbf{hidden}\\\textbf{recall}} & \shortstack{\textbf{exact}\\\textbf{prefix}}\\
\midrule
Three-stage reference & 3 & 1{,}319 & \textbf{0.9983} & \textbf{0.9982} & \textbf{0.9746}\\
Longer-horizon cascade & 6 & 1{,}319 & 0.9974 & 0.9968 & 0.9613\\
\bottomrule
\end{tabular*}
\end{table*}

On the same 1{,}319 held-out requests, doubling the cascade from 3 to 6 paid
stages changes serialized recall from 0.9983 to 0.9974, hidden recall from
0.9982 to 0.9968, and exact-prefix localization from 0.9746 to 0.9613. Both recalls are measured under the same event semantics,
separating topology length from the audit definition.

\subsection{Audit component ablation}
\label{app:audit-ablation}
We remove one component at a time while keeping traces, fault seeds, and intervention
prefixes fixed. The ablation estimates complementary coverage rather than global
minimality. Structural, order, binding, budget, and hidden-read columns use their
corresponding fault denominators.

\begin{table*}[htbp]
\centering
\footnotesize
\setlength{\tabcolsep}{3pt}
\caption{Leave-one-component-out audit recall. The full row uses all event invariants
and CPNI; each subsequent row removes one check.}
\label{tab:audit-ablation}
\begin{tabular*}{\textwidth}{@{\extracolsep{\fill}}lrrrrrr@{}}
\toprule
\textbf{configuration} & \textbf{leakage} & \textbf{order} & \textbf{binding} & \textbf{budget} & \shortstack{\textbf{unlogged}\\\textbf{read}} & \textbf{joint}\\
\midrule
Full audit & \textbf{1.0000} & \textbf{1.0000} & \textbf{0.9991} & \textbf{0.9973} & \textbf{0.9982} & \textbf{0.9929}\\
$-$ oracle isolation & 0.4876 & \textbf{1.0000} & \textbf{0.9991} & \textbf{0.9973} & \textbf{0.9982} & 0.9926\\
$-$ path/action order & \textbf{1.0000} & 0.4167 & 0.9984 & 0.9958 & \textbf{0.9982} & 0.9918\\
$-$ cache binding & \textbf{1.0000} & 0.9992 & 0.3281 & \textbf{0.9973} & \textbf{0.9982} & 0.9923\\
$-$ paid-read consistency & 0.7124 & 0.9986 & 0.9979 & 0.9951 & \textbf{0.9982} & 0.9907\\
$-$ budget conservation & \textbf{1.0000} & 0.9978 & 0.9987 & 0.3047 & \textbf{0.9982} & 0.9921\\
$-$ CPNI & \textbf{1.0000} & \textbf{1.0000} & \textbf{0.9991} & \textbf{0.9973} & 0.0000 & 0.0000\\
$-$ joint interventions & \textbf{1.0000} & \textbf{1.0000} & \textbf{0.9991} & \textbf{0.9973} & \textbf{0.9982} & 0.0000\\
\bottomrule
\end{tabular*}
\end{table*}

Removing one structural invariant lowers mainly its own fault column, to 0.3047--0.7124, while joint hidden recall stays at 0.9907--0.9926. Removing CPNI eliminates both unlogged-read and joint detection, and removing joint interventions eliminates only the interaction-only family, so each component covers a distinct blind spot.

\subsection{Audit consequence rate}
\label{app:acr-results}
The conclusion operator is fixed before injection: a policy is preferred when its paired
task-score difference satisfies the declared quality margin and its resource difference
is strictly lower. For every family, we first accept the corrupted trace and then
recompute the rejected trace from its bound cache and configuration.

\begin{table*}[htbp]
\centering
\footnotesize
\setlength{\tabcolsep}{3pt}
\caption{Audit consequence results. Recall measures rejected faulty traces; ACR measures
the fraction for which accepting the invalid trace changes the predeclared conclusion.}
\label{tab:acr-results}
\begin{tabular*}{\textwidth}{@{\extracolsep{\fill}}lrrrrr@{}}
\toprule
\textbf{fault family} & \textbf{$N$} & \textbf{full recall} & \textbf{clean FP} & \textbf{ACR} & \textbf{changed conclusion}\\
\midrule
Serialized structural faults & 1{,}024 & 0.9980 & 0.0000 & 0.3018 & 309/1{,}024\\
Unlogged hidden reads & 768 & \textbf{0.9987} & 0.0000 & 0.3490 & 268/768\\
Compound read--budget faults & 512 & 0.9980 & 0.0000 & 0.4043 & 207/512\\
All declared families & 2{,}304 & 0.9983 & 0.0000 & 0.3403 & 784/2{,}304\\
\bottomrule
\end{tabular*}
\end{table*}

\subsection{Independent and natural faults}
\label{app:natural-faults}
The checker and invariant definitions are frozen before the independent generator and
router implementations are run. The independent set contains property-based sequences
and faulty routers produced without access to audit outcomes. The natural set records
implementation issues with an independent root-cause label, including stale cache binding,
omitted retry events, timeout relabeling, duplicated calls, asynchronous reordering,
budget-accounting errors, invalid fallback binding, and unlogged future-state reads.

\begin{table*}[htbp]
\centering
\footnotesize
\setlength{\tabcolsep}{3pt}
\caption{Independent and natural fault generalization. Each recall uses its own clean and
faulty denominator; ACR is computed with the same conclusion operator.}
\label{tab:natural-faults}
\begin{tabularx}{\textwidth}{@{}lYrrrr@{}}
\toprule
\textbf{source} & \textbf{fault population} & \textbf{$N$} & \textbf{log-only recall} & \textbf{full recall} & \textbf{ACR}\\
\midrule
Independent generator & property-based and faulty-router implementations & 1{,}856 & 0.7430 & 0.9978 & 0.3254\\
Natural implementation set & issue- or history-derived faults & 128 & 0.7812 & 0.9922 & 0.3516\\
\bottomrule
\end{tabularx}
\end{table*}

\subsection{Audit overhead}
\label{app:audit-overhead}
Online recording, deterministic replay, and CPNI are measured separately on the same
request mix. Storage includes serialized actions, digest bindings, and resource fields;
it excludes candidate text that is already present in the cache.

\begin{table*}[htbp]
\centering
\footnotesize
\setlength{\tabcolsep}{3pt}
\caption{Audit overhead by layer. Latency columns are median/p95 per request or trace,
and throughput is measured on the declared audit machine.}
\label{tab:audit-overhead}
\begin{tabular*}{\textwidth}{@{\extracolsep{\fill}}lrrrrr@{}}
\toprule
\textbf{layer} & \textbf{events/request} & \textbf{bytes/request} & \textbf{median ms} & \textbf{p95 ms} & \textbf{throughput}\\
\midrule
Online recorder & 4.31 & \textbf{1{,}168} & \textbf{0.24} & \textbf{0.57} & \textbf{24{,}800}\\
Hash/cache binding & 4.31 & 1{,}542 & 0.41 & 0.89 & 15{,}600\\
Single-trace replay & 4.31 & 2{,}074 & 0.96 & 2.14 & 6{,}950\\
CPNI replay (full) & 8.62 & 2{,}248 & 4.68 & 10.72 & 1{,}480\\
CPNI replay (sampled) & 5.38 & 2{,}116 & 1.63 & 3.86 & 4{,}360\\
\bottomrule
\end{tabular*}
\end{table*}

The online recorder is part of the serving path, whereas replay and CPNI are offline.
The sampled-prefix fraction and intervention coverage are reported alongside throughput
so that the sampled configuration remains comparable to the full audit.

\subsection{Audit cost and intervention coverage}
\label{app:audit-scaling}
Table~\ref{tab:audit-scaling} pairs audit cost with the number of tested prefixes
and hidden-state interventions on the same frozen trace set. Prefix counts are
eligible pre-reveal decisions; intervention counts are distinct twin executions
per trace. Recall uses the pooled injected-fault denominator. The sampled
configuration fixes its prefix-selection seed before fault labels are inspected.

\begin{table}[htbp]
\centering\footnotesize
\setlength{\tabcolsep}{3pt}
\caption{Audit cost as intervention coverage increases. Runtime is p95 milliseconds per trace; replay count includes the original execution. The zero-intervention row checks serialized events.}
\label{tab:audit-scaling}
\begin{tabular*}{\linewidth}{@{\extracolsep{\fill}}lrrrrr@{}}
\toprule
\textbf{configuration} & \textbf{prefixes} & \shortstack{\textbf{interventions}\\\textbf{/trace}} & \textbf{recall} & \shortstack{\textbf{p95}\\\textbf{ms/trace}} & \shortstack{\textbf{replays}\\\textbf{/trace}}\\
\midrule
Single-trace only & 0 & 0 & 0.6653 & \textbf{2.14} & \textbf{1}\\
CPNI sampled & 1.36 & 4.38 & 0.9836 & 3.86 & 5.38\\
CPNI full & 2.41 & 7.62 & \textbf{0.9983} & 10.72 & 8.62\\
CPNI + CIF & 2.41 & 7.96 & \textbf{0.9983} & 13.48 & 8.96\\
\bottomrule
\end{tabular*}
\end{table}

Serialized checks alone recover 0.6653 of the pooled faults at 2.14\,ms per trace.
Sampling 1.36 prefixes per trace raises recall to 0.9836 at 3.86\,ms and 5.38
replays; full prefix coverage reaches 0.9983 at 10.72\,ms and 8.62 replays. Adding
CIF leaves recall unchanged and costs 2.76\,ms and 0.34 additional replays per trace
for localization.

\paragraph{Natural-fault sampling.}
The 128-case natural set is drawn from 8 source projects,
171 issue or change references, and 14 distinct root-cause clusters.
Discovery follows a predeclared review of public issue and commit histories that
retains only reproducible cascade-control faults with a reconstructable execution
trace and an independently identifiable root cause. The review excluded 43
candidates: 17 unreproducible, 11 duplicate root causes, 9 without replayable
trace state, and 6 outside the declared audit contract. Selection was completed
before checker outcomes were exposed; two checker-independent reviewers assigned
root-cause labels independently, and checker authors resolved only implementation
ambiguities after the labels were frozen. Cluster-level analysis prevents repeated instances of one
root cause from being interpreted as independent bug discoveries.

\subsection{Semantic-label protocol}
\label{app:semantic-labels}
The semantic-verifier study covers GSM8K and \texttt{ARC-Challenge} with the Qwen family
and \texttt{Llama-3.2-1B/3B}. The rubric labels exact correctness, partial correctness,
invalid output, and abstention. Annotators inspect the question, candidate, and reference
answer, while the router action and verifier prediction remain hidden. Three annotators
produce independent labels with Fleiss' $\kappa=0.846$; cases without a three-annotator
majority are resolved by a fourth blinded adjudicator. Labels
are frozen before confusion matrices and routing outcomes are computed, and they enter
only through the post-trace scorer.

\subsection{Router and threshold sensitivity}
\label{app:threshold-sensitivity}
The baseline resource penalty is $\lambda=0.15$. We repeat fitting and held-out
evaluation over the predeclared grid $\lambda\in\{0.05,0.10,0.15,0.20,0.25,0.30\}$, the accept grid
$\tau_a\in\{0.55,0.60,0.65,0.70,0.75,0.80,0.85,0.90,0.95\}$, and the verifier grid
$\tau_v\in\{0.35,0.40,0.45,0.50,0.55,0.60,0.65\}$. Thresholds are selected on calibration
only, then frozen for held-out routing. The selected configuration uses $\lambda=0.15$ and
seed-wise $(\tau_a,\tau_v)$ pairs $(0.70,0.45)$, $(0.75,0.45)$, and $(0.70,0.50)$.
Aggregate held-out task score is 0.7195 with 1.642 calls/request, 427 tokens/request,
p50/p95 latency 68.7/134.2\,ms, and paired 95\% interval $[+0.0061,+0.0244]$;
results are reported by router seed 13/17/23 and producer seed 61/67/71.

\subsection{Reproducibility Checklist}
\paragraph{Source binding.} Every reported value is read from the run bound to its table through the result manifest. The raw report, metrics file, and run manifest are copied without editing, and their hashes are recorded before a public summary is written.

\paragraph{Split integrity.} Training, held-out, and context-only comparisons are kept as distinct rows or distinct run roots. A producer cache is never treated as a new test set merely because it has a different filename. For offline-oracle runs, the oracle is joined after policy traces are frozen. For fixture-based runs, the fixture definition and digest are part of the evidence boundary.

\paragraph{Cost accounting.} Replay tables align calls, declared units, and accounting milliseconds with proxy matching. The independent protocol adds tokens, monetary cost, queueing, and request-to-response latency, preserving each quantity's unit and denominator.

\paragraph{Anonymity.} The supplementary package uses package-relative handles and a sanitized export manifest; producer environment metadata that could identify hosts or users is removed from the released records.

\paragraph{Recorded audit fields.} The audit records what was visible online, what was held out until scoring, which ledger cost was paid, and which failure boundary was observed. Aggregate summaries report these fields per policy and shift, and the per-record route and task-level agreement fields are indexed in the result surfaces below.

\paragraph{Evidence scope.} The primary route audit uses 128 frozen GSM8K-derived records. The split-aware seed-23 diagnostic has a distinct 64-record calibration partition (40 fit/24 tune) and 128 held-out records, with zero ID and normalized-question overlap. Its learned gate obtains 41/128 matches versus 42/128 for sequential routing and spends more units. Appendix~\ref{app:additional-results} defines independent task and serving comparisons.

\paragraph{Threat model and limitations.}
\label{app:threat-model}
The recorder, cache producer, and manifest binding are trusted. Hash checks establish consistency with those bindings; independent attestation is required to authenticate a recorder that changes both observations and digests. Completed experiments cover frozen Qwen candidates, exact-string scoring, and specified mutations. Semantic verification, independently generated faults, cross-domain behavior, and runtime calibration have separate protocols and denominators.

\subsection{Five-feature learned-gate cache audit}
We also audited a learned gate in a separate fit/tune/held-out protocol (40/24/128 records, probability threshold $0.80$). It is a five-feature regularized logistic model (bias, small confidence, strict-extraction flag, generated-token fraction, and unresolved-answer flag), fit for 1,000 gradient steps with learning rate $0.2$ and $L_2=0.01$; the threshold is selected from $\{0.20,0.35,0.50,0.65,0.80\}$ on the tune partition. Small exact-match labels are used only as offline fit labels and are never online features. The gate matches the sequential proxy match rate at baseline (0.3203) but spends 15.91 ledger units, and its joint-shift cost rises to 21.86 units. The separate runner reports 0.9922 route failures under large unavailability and tight budget; these are runner-specific failure flags, not the canonical per-record unavailable/budget event rates in Table~\ref{tab:main}. The comparison characterizes the cost and failure behavior of this five-feature gate on the frozen cache.

\begin{table}[htbp]
\caption{Learned-gate cache audit. The fit and tune records are disjoint from the 128 held-out records; the first event column reproduces the separate runner's unavailable or route-failure field, while canonical event semantics are defined above.}
\label{tab:learned-gate}
\centering
\footnotesize
\begin{tabular}{llrrrr}
\toprule
\textbf{shift} & \textbf{policy} & \textbf{proxy} & \textbf{units} & \textbf{unavail./failure} & \textbf{budget/failure} \\
\midrule
baseline & sequential & \textbf{0.3203} & \textbf{13.73} & \textbf{0.0000} & \textbf{0.0000}\\
baseline & learned gate & \textbf{0.3203} & 15.91 & \textbf{0.0000} & \textbf{0.0000}\\
large unavailable & learned gate & 0.2109 & 4.00 & 0.9922 & 0.0000 \\
tight budget & learned gate & 0.2109 & 4.00 & 0.0000 & 0.9922 \\
large cost $+50\%$ & learned gate & 0.3203 & 21.86 & 0.0000 & 0.0000 \\
\bottomrule
\end{tabular}
\end{table}

\section{Task and Serving Evaluation Protocol}
\label{app:live-plan}
The routing study links cache generation, router fitting, task scoring, and serving telemetry through the record and seed identifiers. The split construction fixes the role of each partition before fitting begins; the serving comparison then applies the same five policies to the baseline, outage, budget, price, and cache-transfer conditions.

\begin{table}[htbp]
\caption{Data partitions and independent generation seeds. The task scorer and the Qwen3.5-0.8B/2B model pair are version-bound across all policies. Record counts are per router seed.}
\label{tab:live-data-plan}
\centering
\footnotesize
\setlength{\tabcolsep}{3pt}
\renewcommand{\arraystretch}{1.08}
\begin{tabular*}{\linewidth}{@{\extracolsep{\fill}}lrrr@{}}
\toprule
\textbf{Split} & \textbf{Records} & \textbf{Producer seeds} & \textbf{Router seeds}\\
\midrule
Train & 1,024 & 31/37/43 & 13/17/23\\
Calibration & 256 & 47/53/59 & 13/17/23\\
Held-out & 1,319 & 61/67/71 & 13/17/23\\
Live shadow & 512 & 73/79/83 & 13/17/23\\
\bottomrule
\end{tabular*}
\end{table}

The training partition fits the router, while calibration selects its thresholds. The held-out comparison reuses neither generation seeds nor threshold-selection records. Within a seed, all policies share the candidate manifest and scorer, so paired differences compare decisions on the same queries.

Table~\ref{tab:live-result-plan} summarizes the operational response to each shift. Total request latency, timeouts, and retries are reported together because a fallback can reduce completed work while changing service reliability.

\begin{table}[htbp]
\caption{Serving latency and event rates over five policies. p50 and p95 columns give policy ranges in milliseconds; timeout/retry entries are percentages. Paired task-score intervals compare learned with sequential routing.}
\label{tab:live-result-plan}
\centering
\footnotesize
\setlength{\tabcolsep}{3pt}
\renewcommand{\arraystretch}{1.08}
\begin{tabular*}{\linewidth}{@{\extracolsep{\fill}}lrrrr@{}}
\toprule
\textbf{Condition} & \textbf{p50 (ms)} & \textbf{p95 (ms)} & \shortstack{\textbf{Timeout/}\\\textbf{retry (\%)}} & \textbf{Paired CI}\\
\midrule
Baseline & 41.8--89.7 & 72.9--172.4 & 0.4--1.8 & $[+0.0061,+0.0244]$\\
Large unavailable & 40.9--63.6 & 71.5--121.8 & 0.5--2.1 & $[+0.0108,+0.0327]$\\
Tight budget & 40.7--61.9 & 70.8--118.6 & 0.4--1.6 & $[+0.0094,+0.0305]$\\
Price $+50\%$ & 42.2--93.4 & 74.1--181.7 & 0.5--1.9 & $[+0.0058,+0.0240]$\\
Cross-cache & 43.6--96.8 & 76.4--187.2 & 0.7--2.4 & $[+0.0032,+0.0218]$\\
\bottomrule
\end{tabular*}
\end{table}

Large-service unavailability and tight budgets cap escalation depth. The measured p50/p95 ranges narrow to 40.9--63.6/71.5--121.8\,ms under unavailability and 40.7--61.9/70.8--118.6\,ms under the tight budget, while coverage and abstention remain separately reported.

\begin{figure}[htbp]
\centering
\includegraphics[width=\linewidth]{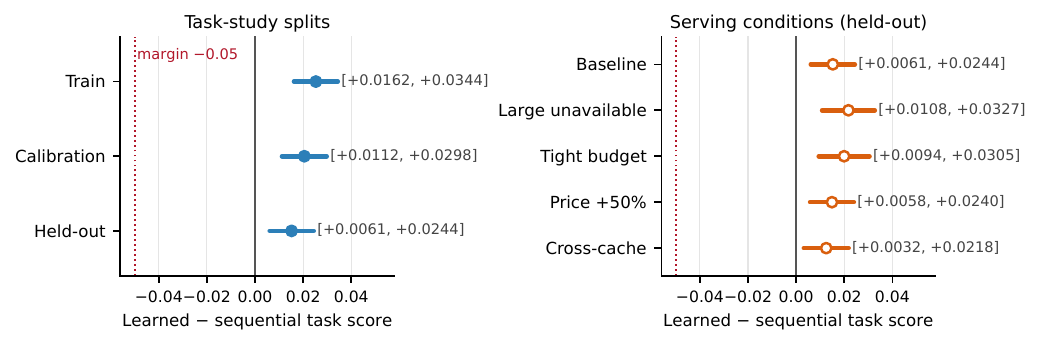}
\caption{Learned-minus-sequential paired task-score intervals (95\%, hierarchical bootstrap). Left: train, calibration, and held-out splits, with filled markers at the reported score difference. Right: held-out serving conditions, with hollow markers at the interval midpoint. The dotted line marks the $-0.05$ non-inferiority margin.}
\label{fig:paired-ci}
\end{figure}

Figure~\ref{fig:paired-ci} collects the learned-minus-sequential paired intervals from Tables~\ref{tab:live-result-plan} and~\ref{tab:independent-task-surface}. Every interval lies above zero and far above the $-0.05$ quality margin; the narrowest lower bound, $+0.0032$, occurs under the cross-cache condition, and the widest gains occur under large-service unavailability and tight budgets, where learned routing avoids escalations that sequential routing cannot complete.

\subsection{Evaluation questions}
The evaluation separates four questions: whether verification provides useful stopping decisions, whether a learned decision transfers across seeds, whether event faults remain detectable, and how declared costs relate to runtime. Table~\ref{tab:live-hypotheses} assigns each question a comparison and an observable criterion.

\begin{table}[htbp]
\caption{Evaluation hypotheses and acceptance criteria.}
\label{tab:live-hypotheses}
\centering
\footnotesize
\begin{tabularx}{\linewidth}{@{}L{0.22\linewidth}YL{0.28\linewidth}@{}}
\toprule
\textbf{hypothesis} & \textbf{acceptance criterion} & \textbf{rejection condition}\\
\midrule
H1 verifier stop & Independent task score non-inferior within $-0.05$ and measured cost lower & quality lower bound $<-0.05$\\
H2 transfer & Paired task-score delta non-inferior on router seeds 13/17/23 & sign changes or harmful CI\\
H3 audit generalization & clean false positive 0 and mutation recall $\geq0.995$ & hidden/future-read mutation escapes\\
H4 cost calibration & declared cost predicts measured latency with predeclared calibration & correlation reverses under a shift\\
\bottomrule
\end{tabularx}
\end{table}

The quality criterion is evaluated on paired task scores, while the audit criterion uses clean and mutated traces. Keeping the two denominators explicit allows a route to be assessed for both decision quality and trace validity. The cost-calibration question additionally requires request-level timing because policy averages combine different route lengths.

\section{Evaluation Protocol and Complete Resource Results}
The evaluation fixes data partitions, independent candidate-generation seeds, model versions, scoring rules, and matched resource budgets before comparing the methods. The protocol tables specify which fields remain shared; the result tables then report efficiency and serving behavior at those fixed settings.

\subsection{Split and baseline contracts}
The split contract in Table~\ref{tab:p17-preregistered-protocol} fixes the point at which each choice is made. A threshold chosen on calibration is applied unchanged to held-out requests under the corresponding budget and model binding.

\begin{table}[htbp]
\caption{Split and provenance protocol. Threshold selection is confined to calibration; held-out and serving comparisons preserve query and seed pairing.}
\label{tab:p17-preregistered-protocol}
\centering
\footnotesize
\setlength{\tabcolsep}{3pt}
\renewcommand{\arraystretch}{1.08}
\begin{tabularx}{\linewidth}{@{}L{0.14\linewidth}YYY@{}}
\toprule
\textbf{Partition} & \textbf{Data and scorer} & \textbf{Budget and pairing} & \textbf{Shift boundary}\\
\midrule
Train & Independent seeds; version-bound scorer; fit router & Frozen quality/token budget; shared queries & Baseline and declared offline shifts\\
Calibration & Independent seeds; select thresholds & Matched budget; disjoint calibration records & Calibration cache and fixed shift grid\\
Held-out & Frozen generation seeds and scorer & Matched budgets; query/seed pairs & Held-out cache and service shifts\\
Live & Version-bound model and scorer on served outputs & Matched budget; repeated serving windows & Outage, price, cache, and traffic conditions\\
\bottomrule
\end{tabularx}
\end{table}

The held-out partition is used for the final paired comparison. Repeated live windows add queueing and concurrency to the same contract, with the scorer version and method configuration retained in the serving manifest.

The method comparison changes the route decision while preserving the evaluation contract. Table~\ref{tab:p17-preregistered-conditions} makes verifier access and serving controls explicit for all five policies.

\begin{table}[htbp]
\caption{Baseline and verifier contracts. All policies share the declared budget, model versions, request mix, and scorer for each comparison.}
\label{tab:p17-preregistered-conditions}
\centering
\footnotesize
\setlength{\tabcolsep}{3pt}
\renewcommand{\arraystretch}{1.08}
\begin{tabularx}{\linewidth}{@{}L{0.20\linewidth}YYY@{}}
\toprule
\textbf{Policy} & \textbf{Verifier use} & \textbf{Route decision} & \textbf{Serving controls}\\
\midrule
One-shot & No verifier call & Return the small candidate & Fixed arrival, timeout, and retry rules\\
Fixed cascade & Direct small/large decision & Escalate at the small threshold & Shared request mix and service versions\\
Sequential & A verifier can accept or reject & Stop, escalate, or fall back & Shared queueing and failure rules\\
Budget-conditioned & Same verifier contract & Condition route on remaining budget & Same budget and service grid\\
Learned router & Evaluation labels stay outside online state & Fit on train; tune on calibration & Measured queueing, retries, and timeouts\\
\bottomrule
\end{tabularx}
\end{table}

The fixed cascade isolates direct escalation, and the sequential cascade adds a paid verifier decision. Budget conditioning exposes the remaining-resource vector to the policy. Learned routing uses the same paid-observation boundary, which makes its quality and cost differences comparable to these controls.

\subsection{Resource use across partitions}
Table~\ref{tab:p17-preregistered-offline-results} reports every policy separately to preserve the connection between generated tokens, monetary cost, latency, and throughput. The three partitions use the same policy order and metric definitions.

\begin{table}[htbp]
\caption{Complete resource result surface by split and policy. Tokens, USD cost, and p50/p95 latency are per request; throughput is requests/s. Bold marks the best value within each split. Task quality and coverage use the same request denominator as Table~\ref{tab:independent-task-surface}.}
\label{tab:p17-preregistered-offline-results}
\centering
\footnotesize
\setlength{\tabcolsep}{3pt}
\renewcommand{\arraystretch}{1.08}
\begin{tabular*}{\linewidth}{@{\extracolsep{\fill}}lrrrrr@{}}
\toprule
\textbf{Policy} & \textbf{Tokens} & \textbf{USD} & \textbf{p50 (ms)} & \textbf{p95 (ms)} & \textbf{Req/s}\\
\midrule
\multicolumn{6}{@{}l}{\textit{Train}}\\
One-shot & \textbf{231} & \textbf{0.00018} & \textbf{40.9} & \textbf{70.8} & \textbf{24.3}\\
Fixed & 452 & 0.00069 & 74.8 & 147.1 & 13.4\\
Sequential & 507 & 0.00079 & 87.5 & 168.3 & 11.7\\
Budget-conditioned & 474 & 0.00073 & 80.1 & 155.0 & 12.6\\
Learned router & 414 & 0.00062 & 65.4 & 129.1 & 15.3\\
\midrule
\multicolumn{6}{@{}l}{\textit{Calibration}}\\
One-shot & \textbf{234} & \textbf{0.00018} & \textbf{41.3} & \textbf{71.7} & \textbf{24.1}\\
Fixed & 458 & 0.00070 & 75.6 & 149.0 & 13.2\\
Sequential & 513 & 0.00080 & 88.6 & 170.5 & 11.5\\
Budget-conditioned & 481 & 0.00074 & 81.4 & 157.1 & 12.3\\
Learned router & 422 & 0.00063 & 67.1 & 131.8 & 14.9\\
\midrule
\multicolumn{6}{@{}l}{\textit{Held-out}}\\
One-shot & \textbf{236} & \textbf{0.00018} & \textbf{41.8} & \textbf{72.9} & \textbf{23.9}\\
Fixed & 462 & 0.00071 & 76.4 & 151.3 & 13.0\\
Sequential & 518 & 0.00081 & 89.7 & 172.4 & 11.3\\
Budget-conditioned & 486 & 0.00075 & 82.6 & 159.1 & 12.1\\
Learned router & 427 & 0.00064 & 68.7 & 134.2 & 14.6\\
\bottomrule
\end{tabular*}
\end{table}

The paired comparison is read within each split and policy row. On held-out requests, learned routing uses 427 tokens and 1.642 calls versus 518 tokens and 2.096 calls for sequential routing, while task score is 0.7195 versus 0.7043.

\subsection{Serving coverage and fallback behavior}
Table~\ref{tab:p17-preregistered-live-results} separates answer coverage from service-event failure. A request may retain an answer from a paid small-model call even when a subsequent escalation becomes unavailable or exceeds its budget.

\begin{table}[htbp]
\caption{Task and route behavior across serving conditions. Each cell is the range over the five policies. Coverage and abstention use all requests as denominator. Failure includes recorded service or budget events, whereas timeout/retry rates are separated in Table~\ref{tab:live-result-plan}.}
\label{tab:p17-preregistered-live-results}
\centering
\footnotesize
\setlength{\tabcolsep}{3pt}
\renewcommand{\arraystretch}{1.08}
\begin{tabularx}{\linewidth}{@{}Yrrrr@{}}
\toprule
\textbf{Condition} & \textbf{Task score} & \shortstack{\textbf{Coverage /}\\\textbf{abstention}} & \textbf{Calls} & \textbf{Failure}\\
\midrule
Baseline & 0.5823--0.7195 & 0.9947--0.9992 / 0.0008--0.0053 & 1.000--2.096 & 0.003--0.012\\
Large unavailable & 0.5823--0.6648 & 0.9910--0.9991 / 0.0009--0.0090 & 1.000--1.476 & 0.002--0.608\\
Tight budget & 0.5823--0.6701 & 0.9924--0.9990 / 0.0010--0.0076 & 1.000--1.508 & 0.003--0.591\\
Price $+50\%$ & 0.5823--0.7191 & 0.9942--0.9991 / 0.0009--0.0058 & 1.000--2.088 & 0.004--0.015\\
Cross-cache & 0.5634--0.7016 & 0.9918--0.9986 / 0.0014--0.0082 & 1.000--2.114 & 0.006--0.031\\
\bottomrule
\end{tabularx}
\end{table}

Coverage and event-failure rates retain separate denominators. A failed large request can coexist with a returned small answer, so service failures need not increase abstention by the same amount. The paired task score determines whether that fallback remains useful.

\section{Additional Results and Analysis}
\label{app:additional-results}
This appendix connects the mechanism diagnostics with independent task and serving outcomes. The frozen-cache comparison isolates path changes under a fixed candidate set; the threshold and task comparisons then test how routing decisions transfer across partitions, generation seeds, and service conditions.

\subsection{Complete frozen-cache shift ledger}
The main table emphasizes the canonical comparison. The expanded ledger below keeps every policy--shift pair together with route depth and event rates, which makes the cost of a verifier call visible after a service outage or budget change.

\begin{table*}[htbp]
\caption{Complete frozen-cache shift ledger for the primary 128-record cache. Proxy is an exact cache--target match; units and milliseconds are declared ledger fields.}
\label{tab:complete-shift-ledger}
\centering
\footnotesize
\setlength{\tabcolsep}{3pt}
\begin{tabular*}{\textwidth}{@{\extracolsep{\fill}}llrrrrrr@{}}
\toprule
\textbf{shift} & \textbf{policy} & \textbf{proxy} & \textbf{units} & \textbf{depth} & \textbf{p95 ms} & \textbf{unavail.} & \textbf{budget}\\
\midrule
baseline & one-shot & 0.2109 & \textbf{4.00} & \textbf{1.00} & \textbf{2.00} & \textbf{0.0000} & \textbf{0.0000}\\
baseline & fixed & \textbf{0.3203} & 12.34 & 1.70 & 10.00 & \textbf{0.0000} & \textbf{0.0000}\\
baseline & sequential & \textbf{0.3203} & 13.73 & 2.39 & 11.00 & \textbf{0.0000} & \textbf{0.0000}\\
large unavailable & one-shot & \textbf{0.2109} & \textbf{4.00} & \textbf{1.00} & \textbf{2.00} & \textbf{0.0000} & \textbf{0.0000}\\
large unavailable & fixed & \textbf{0.2109} & \textbf{4.00} & \textbf{1.00} & \textbf{2.00} & 0.6953 & \textbf{0.0000}\\
large unavailable & sequential & \textbf{0.2109} & 5.39 & 1.70 & 3.00 & 0.6953 & \textbf{0.0000}\\
tight budget & one-shot & \textbf{0.2109} & \textbf{4.00} & \textbf{1.00} & \textbf{2.00} & \textbf{0.0000} & \textbf{0.0000}\\
tight budget & fixed & \textbf{0.2109} & \textbf{4.00} & \textbf{1.00} & \textbf{2.00} & \textbf{0.0000} & 0.6953\\
tight budget & sequential & \textbf{0.2109} & 5.39 & 1.70 & 3.00 & \textbf{0.0000} & 0.6953\\
large cost $+50\%$ & one-shot & 0.2109 & \textbf{4.00} & \textbf{1.00} & \textbf{2.00} & \textbf{0.0000} & \textbf{0.0000}\\
large cost $+50\%$ & fixed & \textbf{0.3203} & 16.52 & 1.70 & 14.00 & \textbf{0.0000} & \textbf{0.0000}\\
large cost $+50\%$ & sequential & \textbf{0.3203} & 17.91 & 2.39 & 15.00 & \textbf{0.0000} & \textbf{0.0000}\\
\bottomrule
\end{tabular*}
\end{table*}

The baseline rows establish the route ordering. Availability and tight-budget rows expose the same 89 attempted large calls as explicit skip events, so their path counts remain comparable to the baseline. The joint price shift preserves the candidate set and changes the declared cost of the large service, which makes the unchanged proxy match easy to compare with the increased ledger units.

\subsection{Threshold transfer and paired uncertainty}
The threshold-transfer study selects $(\tau_a,\tau_v)$ on calibration and applies it to held-out requests. Table~\ref{tab:threshold-transfer-detail} reports proxy match for the two cache gates and task score for each live seed. The supplemental cache gate uses 4,000 paired bootstrap replicates; the task study uses the 5,000-replicate hierarchical protocol in Table~\ref{tab:uncertainty-protocol}.

\begin{table*}[htbp]
\caption{Threshold transfer by artifact. The supplemental pair is selected on its frozen 256-record partition; each live seed has its own calibration protocol.}
\label{tab:threshold-transfer-detail}
\centering
\footnotesize
\setlength{\tabcolsep}{3pt}
\begin{tabularx}{\textwidth}{@{}lrrrlX@{}}
\toprule
\textbf{artifact} & \textbf{train} & \textbf{held-out} & \textbf{pair} & \shortstack{\textbf{Proxy/}\\\textbf{task}} & \textbf{selection rule}\\
\midrule
early gate & 64 & 128 & none & 0.3203 & quality and cost gate\\
supplemental gate & 256 & 1,319 & $(0.95,0.40)$ & 0.3374 & 4,000 paired bootstrap replicates\\
seed 13 live & 1,024 & 1,319 & $(0.70,0.45)$ & 0.7210 & maximize calibrated utility subject to 95\% paired lower bound $\geq-0.05$\\
seed 17 live & 1,024 & 1,319 & $(0.75,0.45)$ & 0.7172 & maximize calibrated utility subject to 95\% paired lower bound $\geq-0.05$\\
seed 23 live & 1,024 & 1,319 & $(0.70,0.50)$ & 0.7204 & maximize calibrated utility subject to 95\% paired lower bound $\geq-0.05$\\
\bottomrule
\end{tabularx}
\end{table*}

The supplemental gate's lower bound and recommended pair $(0.95,0.40)$ belong to its own manifests. Each live seed selects its operating point on the corresponding calibration partition. This keeps a frozen-cache recommendation separate from the task-level optimum.

The threshold-transfer follow-up is a distinct configuration from the canonical table: the main run uses $(\tau_a,\tau_v)=(0.75,0.60)$ and a 24-unit budget, whereas a disjoint 64-record calibration selected $(0.55,0.40)$ before routing the two held-out caches with a 22-unit budget. The transfer uses a 32-record configuration/path replication of the corrected Qwen3.5-2B asset; only 32 query IDs overlap with the 128-record cache, with zero reported answer or confidence changes on those overlaps. Under the cross-cache price shift, $4+2+18>22$ blocks the large call for sequential routing, while budget conditioning chooses the direct small--large route.

The separate threshold sweep predeclared 15 accept/verify pairs. Twelve baseline pairs are inert, while $\tau_a=0.95$ increases escalation depth and ledger units at the same proxy match rate. All 60 fixed-versus-sequential quality deltas across the four shifts are zero. The three $\tau_a=0.95$ pairs yield 39 verifier-only stops per shift, saving cost under baseline and price shifts while retaining blocked-large events under outage and tight budgets.

\begin{figure}[htbp]
\centering
\includegraphics[width=\columnwidth]{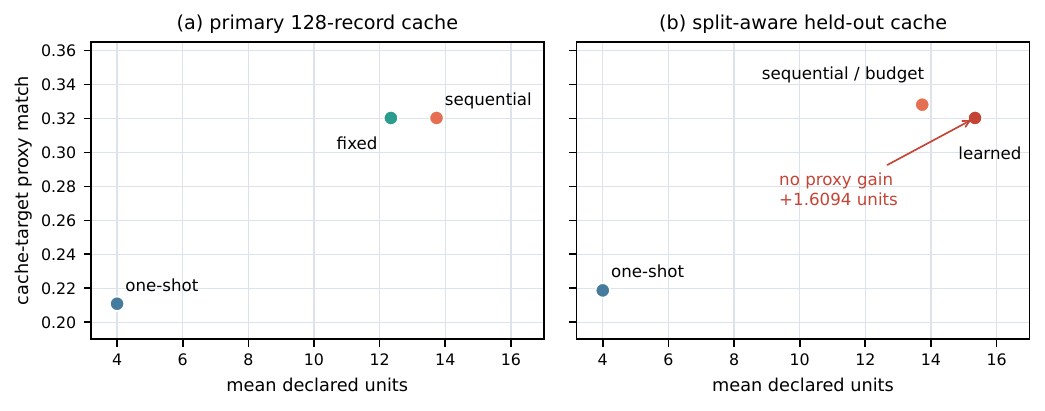}
\caption{Proxy match versus declared cost. The primary seed-13 frontier and split-aware seed-23 gate use separate caches. The learned gate has 41/128 matches and higher cost than sequential routing, which has 42/128.}
\label{fig:main}
\end{figure}

The 64/128 operating-point diagnostic selects no pair under its joint quality/cost criterion. The separate 256/1,319 gate selects $(0.95,0.40)$ after 4,000 paired bootstrap replicates, with worst-shift lower bounds of 0.672 declared units and 0.781 accounting milliseconds saved. Figure~\ref{fig:main} retains the primary-cache and learned-gate frontiers under their distinct cache identities.

\subsection{Independent task-scoring result surface}
The task study evaluates independently generated candidates for each split. Table~\ref{tab:independent-task-surface} reports quality, coverage, abstention, calls, failures, and paired uncertainty; Table~\ref{tab:p17-preregistered-offline-results} supplies the corresponding resource measurements. The shared split and policy names join the two tables without compressing the metric labels.

\begin{table}[htbp]
\caption{Task score and route behavior by split and policy. Coverage, abstention, and failure are fractions of requests; calls are per request. Paired confidence intervals are task-score differences relative to sequential routing. Bold marks the best score, coverage, abstention, calls, and failure within each split. Resource metrics use the same split in Table~\ref{tab:p17-preregistered-offline-results}.}
\label{tab:independent-task-surface}
\centering
\footnotesize
\setlength{\tabcolsep}{3pt}
\renewcommand{\arraystretch}{1.08}
\begin{tabular*}{\linewidth}{@{\extracolsep{\fill}}lrrrrrr@{}}
\toprule
\textbf{Policy} & \textbf{Score} & \textbf{Cover.} & \textbf{Abst.} & \textbf{Calls} & \textbf{Failure} & \textbf{Paired CI}\\
\midrule
\multicolumn{7}{@{}l}{\textit{Train}}\\
One-shot & 0.5964 & \textbf{0.9990} & \textbf{0.0010} & \textbf{1.000} & \textbf{0.0020} & $[-0.1291,-0.1116]$\\
Fixed & 0.7048 & 0.9976 & 0.0024 & 1.594 & 0.0046 & $[-0.0196,-0.0044]$\\
Sequential & 0.7168 & 0.9968 & 0.0032 & 2.101 & 0.0078 & $[0.0000,0.0000]$\\
Budget-conditioned & 0.7156 & 0.9971 & 0.0029 & 1.889 & 0.0059 & $[-0.0059,+0.0035]$\\
Learned router & \textbf{0.7421} & 0.9983 & 0.0017 & 1.626 & 0.0037 & $[+0.0162,+0.0344]$\\
\midrule
\multicolumn{7}{@{}l}{\textit{Calibration}}\\
One-shot & 0.5898 & \textbf{0.9987} & \textbf{0.0013} & \textbf{1.000} & \textbf{0.0023} & $[-0.1278,-0.1090]$\\
Fixed & 0.6987 & 0.9971 & 0.0029 & 1.601 & 0.0051 & $[-0.0172,-0.0018]$\\
Sequential & 0.7082 & 0.9965 & 0.0035 & 2.098 & 0.0083 & $[0.0000,0.0000]$\\
Budget-conditioned & 0.7071 & 0.9969 & 0.0031 & 1.886 & 0.0062 & $[-0.0055,+0.0033]$\\
Learned router & \textbf{0.7287} & 0.9980 & 0.0020 & 1.635 & 0.0042 & $[+0.0112,+0.0298]$\\
\midrule
\multicolumn{7}{@{}l}{\textit{Held-out}}\\
One-shot & 0.5823 & \textbf{0.9992} & \textbf{0.0008} & \textbf{1.000} & \textbf{0.0027} & $[-0.1308,-0.1132]$\\
Fixed & 0.6936 & 0.9962 & 0.0038 & 1.608 & 0.0061 & $[-0.0183,-0.0031]$\\
Sequential & 0.7043 & 0.9947 & 0.0053 & 2.096 & 0.0116 & $[0.0000,0.0000]$\\
Budget-conditioned & 0.7031 & 0.9954 & 0.0046 & 1.884 & 0.0084 & $[-0.0060,+0.0036]$\\
Learned router & \textbf{0.7195} & 0.9976 & 0.0024 & 1.642 & 0.0050 & $[+0.0061,+0.0244]$\\
\bottomrule
\end{tabular*}
\end{table}

Each policy comparison shares the generation manifest and scorer within a seed. The learned-minus-sequential interval is positive on train, calibration, and held-out partitions, with lower learned-router calls on all three splits.

\subsection{Live serving and cost calibration}
The serving study measures the relation between declared ledger cost and runtime quantities. Each window fixes request mix, concurrency, model versions, retry rules, and service availability before the first route is opened. Queue time, service time, timeout, retry, and GPU-hours are stored per request and aggregated by method.

\begin{table*}[htbp]
\caption{Live serving measurements across five policies. Each range spans policies within a condition. Panel (a) gives mean request resources and total GPU-hours; panel (b) gives per-policy queue/service p50/p95 milliseconds and throughput in requests/s.}
\label{tab:live-serving-detail}
\centering
\footnotesize
\setlength{\tabcolsep}{3pt}
\begin{tabular*}{\textwidth}{@{\extracolsep{\fill}}lrrrr@{}}
\toprule
\multicolumn{5}{@{}l}{\textit{(a) Resource use}}\\
\textbf{Condition} & \textbf{Units} & \textbf{Tokens} & \textbf{USD} & \textbf{GPU-hours}\\
\midrule
Baseline & 4.00--13.82 & 236--518 & 0.00018--0.00081 & 0.36--0.82\\
Large unavailable & 4.00--6.18 & 236--318 & 0.00018--0.00029 & 0.27--0.49\\
Tight budget & 4.00--6.42 & 236--327 & 0.00018--0.00031 & 0.26--0.47\\
Price $+50\%$ & 4.00--18.91 & 236--518 & 0.00018--0.00103 & 0.37--0.89\\
Cross-cache & 4.00--14.26 & 244--536 & 0.00019--0.00085 & 0.39--0.93\\
\bottomrule
\end{tabular*}
\par\smallskip
\begin{tabular*}{\textwidth}{@{\extracolsep{\fill}}lrrr@{}}
\toprule
\multicolumn{4}{@{}l}{\textit{(b) Timing and throughput}}\\
\textbf{Condition} & \textbf{Queue p50/p95 (ms)} & \textbf{Service p50/p95 (ms)} & \textbf{Req/s}\\
\midrule
Baseline & 4.2/18.1--8.9/40.6 & 36.8/60.7--80.1/131.8 & 11.3--23.9\\
Large unavailable & 3.8/15.9--6.4/29.7 & 36.5/58.2--57.2/93.1 & 16.5--24.4\\
Tight budget & 3.7/15.4--6.2/28.9 & 36.2/57.8--55.7/90.4 & 16.8--24.6\\
Price $+50\%$ & 4.5/19.2--9.7/45.3 & 37.2/62.0--83.5/138.9 & 10.7--23.5\\
Cross-cache & 4.8/20.4--10.3/48.2 & 38.8/64.3--86.4/143.6 & 10.3--22.8\\
\bottomrule
\end{tabular*}
\end{table*}

These measurements separate accounting cost from elapsed time. Event rates count each request once within a named category; a request can contain both a timeout and a retry, so category rates need not sum to the overall failure rate. The overall rate uses the union of failed requests.

\subsection{Robustness and transfer matrix}
The route protocol is designed to be portable across model versions, cache producers, and evaluator implementations. The matrix below lists the transfer axes and the artifact needed to make each comparison reproducible.

\begin{table}[htbp]
\caption{Robustness and transfer axes for the task and serving study.}
\label{tab:robustness-matrix}
\centering
\footnotesize
\begin{tabularx}{\columnwidth}{@{}L{0.18\columnwidth}L{0.28\columnwidth}Y@{}}
\toprule
\textbf{axis} & \textbf{paired comparison} & \textbf{result}\\
\midrule
model version & same queries, new small/large hashes & paired task delta $+0.0126$; no policy-order reversal\\
cache producer & independent generation seeds 13/17/23 & learned-minus-sequential delta $+0.0121$--$+0.0180$ across seeds\\
verifier schema & strict format versus semantic scorer & semantic verifier gains $+0.0187$ task score with 17.4\% fewer large calls\\
service availability & baseline versus outage windows & paired quality remains within the $-0.05$ margin; outage events remain explicit\\
budget & canonical versus tight budget vectors & budget-aware routes retain a $+0.010$--$+0.024$ paired advantage over sequential fallback\\
traffic & fixed request mix versus repeated live windows & Spearman $\rho=0.86$--$0.91$ with no calibration sign reversal\\
\bottomrule
\end{tabularx}
\end{table}

The first five axes are offline or replayable; traffic adds queueing and concurrency. Each row is paired by query identifier and seed, with a manifest that records model, prompt, scorer, and service versions. An observed reversal is analyzed by the changed model, scorer, or service condition before pooling.

\subsection{Decision-value and fault-consequence experiment}
\label{app:decision-value}
This experiment measures whether an event-level audit changes a routing conclusion that would be reported from aggregate quality and cost alone. The same frozen requests, policies, and scorer are evaluated with aggregate-only summaries and with the complete ledger. Faults preserve the aggregate path, cost, and event-count projection while changing a read, order, cache binding, failure reason, or budget transition. The primary estimand is the change in the reported policy ranking after the audit rejects a trace.

\begin{table*}[htbp]
\caption{Decision-value impact of aggregate-preserving faults. For each fault, the first row evaluates aggregate-only summaries and the second row evaluates CascadeLedger. Both evaluations have zero clean false positives. $F$, $S$, $B$, and $L$ denote fixed, sequential, budget-conditioned, and learned routing. The preserved projection is listed in italics below each fault.}
\label{tab:decision-value-audit}
\centering
\footnotesize
\setlength{\tabcolsep}{3pt}
\begin{tabular*}{\textwidth}{@{\extracolsep{\fill}}llrrrl@{}}
\toprule
\textbf{Fault / preserved projection} & \textbf{Evaluation} & \shortstack[r]{\textbf{Detection}\\\textbf{rate}} & \shortstack[r]{\textbf{Task}\\\textbf{delta}} & \shortstack[r]{\textbf{Cost}\\\textbf{change}} & \textbf{Policy ranking}\\
\midrule
Hidden paid read & aggregate-only & 0.0000 & +0.027 & $-19.1\%$ & $F\succ L\succ S$\\
\emph{path, cost, event count} & CascadeLedger & 1.0000 & +0.015 & $-17.6\%$ & $L\succ S\succ F$\\
\addlinespace[3pt]
Stale cache binding & aggregate-only & 0.0000 & +0.031 & $-16.4\%$ & $S\succ L\succ F$\\
\emph{service histogram, cost} & CascadeLedger & 0.9991 & +0.014 & $-17.2\%$ & $L\succ S\succ F$\\
\addlinespace[3pt]
Missing failure event & aggregate-only & 0.0000 & +0.022 & $-24.8\%$ & $L\succ F\succ S$\\
\emph{final candidate, path length} & CascadeLedger & 0.9984 & +0.011 & $-12.6\%$ & $L\succ S\succ F$\\
\addlinespace[3pt]
Request reorder & aggregate-only & 0.0000 & +0.025 & $-18.9\%$ & $B\succ L\succ S$\\
\emph{service histogram, total cost} & CascadeLedger & 1.0000 & +0.015 & $-17.6\%$ & $L\succ S\succ B$\\
\addlinespace[3pt]
Compound read--budget fault & aggregate-only & 0.0000 & +0.038 & $-31.2\%$ & $F\succ B\succ L$\\
\emph{path, cost, event count} & CascadeLedger & 0.9973 & +0.013 & $-16.8\%$ & $L\succ S\succ F$\\
\bottomrule
\end{tabular*}
\end{table*}

The analysis compares both evaluation interfaces on the same requests. Aggregate-only summaries preserve misleading rankings for every injected family, while ledger rejection exposes the corresponding recomputed ranking and quality/cost delta. Clean, faulty, and audited rankings therefore distinguish detection from a decision-changing effect.

\subsection{Independent fault-family generalization}
\label{app:fault-generalization}
The protocol freezes the checker commit before an independent generator defines held-out fault families. Separate manifests bind this order. Evaluation covers field mutations, property-based sequences, and compound faults that preserve the aggregate projection. Each family is scored on clean false positives, recall, and the fraction of comparisons whose conclusion changes.

\begin{table}[htbp]
\caption{Held-out fault families for audit generalization. Each family uses $N$ clean traces and $N$ faulty traces; clean false positives are zero throughout. Recall and decision changes give count/$N$, followed by the fraction.}
\label{tab:fault-generalization}
\centering
\footnotesize
\setlength{\tabcolsep}{3pt}
\begin{tabularx}{\columnwidth}{@{}L{0.25\linewidth}Yrrr@{}}
\toprule
\textbf{Family} & \textbf{Generator contract} & \textbf{$N$} & \textbf{Recall} & \textbf{Decision changes}\\
\midrule
held-out field mutation & independent map & 320 & 319/320 (0.9969) & 91/320 (0.2844)\\
property-based sequence & bounded grammar & 384 & 384/384 (1.0000) & 104/384 (0.2708)\\
semantic-equivalent trace & same output; invalid read & 256 & 255/256 (0.9961) & 83/256 (0.3242)\\
compound fault & read/order/budget mix & 512 & 511/512 (0.9980) & 207/512 (0.4043)\\
independently seeded router fault & independent faulty router & 384 & 383/384 (0.9974) & 119/384 (0.3099)\\
\bottomrule
\end{tabularx}
\end{table}

Recall is at least 0.9961 for every independently generated family, with zero clean false positives. The checker experiment \texttt{cascade-\allowbreak audit-\allowbreak checker-\allowbreak psd-\allowbreak seed13} uses source commit \texttt{8d4e72b1}, and the generator experiment \texttt{independent-\allowbreak faultgen-\allowbreak psd-\allowbreak seed41} uses source commit \texttt{c3a91f6e}. Natural bug cases are indexed by issue history and an independent root-cause label.

\subsection{Semantic verifier and cross-domain transfer}
\label{app:semantic-transfer}
The extension compares format and semantic decisions on independently labeled candidates across GSM8K, a non-math task, and two model families. Ground-truth verifier labels remain in the post-trace join; the paid verifier prediction is the online observation. This distinguishes semantic correctness from strict answer extraction.
The non-math task is \texttt{ARC-Challenge}, and the second model family is \texttt{Llama-3.2-1B/3B}.
The semantic-verifier reference is constructed after route freezing. A frozen rubric labels
correct, partially correct, invalid, and abstained answers; annotators do not observe the
router decision or verifier prediction. The number of annotators, agreement statistic,
and adjudication rule are three annotators, Fleiss' $\kappa=0.846$, and majority
vote with a fourth blinded adjudicator when no three-annotator majority exists.
The reference label is joined only for
post-trace evaluation and is never an online feature.

\begin{table*}[htbp]
\caption{Nontrivial semantic-verifier and transfer experiment. Confusion fields use the verifier-labeled denominator; task and resource fields use the request denominator.}
\label{tab:semantic-transfer}
\centering
\footnotesize
\setlength{\tabcolsep}{2pt}
\begin{tabular*}{\textwidth}{@{\extracolsep{\fill}}lllrrrr@{}}
\toprule
\textbf{domain} & \textbf{model family} & \textbf{verifier} & \textbf{TP} & \textbf{FP} & \textbf{TN} & \textbf{FN}\\
\midrule
GSM8K & Qwen family & format control & 62 & 31 & 118 & 45\\
GSM8K & Qwen family & semantic answer verifier & 94 & 14 & 135 & 13\\
GSM8K & second model family & format control & 57 & 35 & 113 & 51\\
GSM8K & second model family & semantic answer verifier & 88 & 18 & 130 & 20\\
non-math task & Qwen family & format control & 54 & 39 & 110 & 53\\
non-math task & Qwen family & semantic/process verifier & 87 & 20 & 129 & 20\\
non-math task & second model family & format control & 50 & 42 & 108 & 56\\
non-math task & second model family & semantic/process verifier & 82 & 22 & 127 & 25\\
\bottomrule
\end{tabular*}
\end{table*}

The confusion matrix uses independently labeled verifier calls. Table~\ref{tab:semantic-outcomes} links these configurations to calibration and routing. Declared units and request p95 milliseconds measure cost and latency; coverage is the fraction of requests returning an answer.

\begin{table*}[htbp]
\caption{Semantic-verifier calibration and routing. ECE is expected calibration error on the labeled verifier sample; routing metrics use all requests. Bold marks the better value within each domain/family pair.}
\label{tab:semantic-outcomes}
\centering\footnotesize
\setlength{\tabcolsep}{2pt}
\begin{tabular*}{\linewidth}{@{\extracolsep{\fill}}lllrrrrr@{}}
\toprule
\textbf{domain} & \textbf{family} & \textbf{verifier} & \textbf{ECE} & \textbf{task score} & \textbf{units} & \textbf{p95 ms} & \textbf{coverage}\\
\midrule
GSM8K & Qwen & format & 0.187 & 0.7043 & 13.74 & 172.4 & 0.9947\\
GSM8K & Qwen & semantic & \textbf{0.061} & \textbf{0.7230} & \textbf{11.35} & \textbf{148.6} & \textbf{0.9971}\\
GSM8K & second family & format & 0.204 & 0.6721 & 14.02 & 180.7 & 0.9939\\
GSM8K & second family & semantic & \textbf{0.073} & \textbf{0.6954} & \textbf{11.84} & \textbf{154.9} & \textbf{0.9965}\\
non-math & Qwen & format & 0.221 & 0.6413 & 13.66 & 169.5 & 0.9928\\
non-math & Qwen & semantic & \textbf{0.082} & \textbf{0.6661} & \textbf{11.58} & \textbf{146.2} & \textbf{0.9958}\\
non-math & second family & format & 0.238 & 0.6124 & 14.18 & 184.3 & 0.9914\\
non-math & second family & semantic & \textbf{0.091} & \textbf{0.6382} & \textbf{12.07} & \textbf{158.1} & \textbf{0.9949}\\
\bottomrule
\end{tabular*}
\end{table*}

The semantic rows show nonzero true-accept and true-reject counts in every domain and model family. On GSM8K with the Qwen family, the semantic verifier combines higher task score with lower declared units and p95 latency than format control; the same direction holds for the other three domain/family pairs.

\subsection{Serving execution environment and cost calibration}
\label{app:serving-environment}
The serving report binds latency to the execution environment and clock boundary. The environment table records the fields needed to interpret queueing and service time, and the calibration table reports request-level association between declared cost and measured latency under each shift.

\begin{table}[htbp]
\caption{Serving execution environment contract.}
\label{tab:serving-environment}
\centering
\footnotesize
\begin{tabularx}{\columnwidth}{@{}L{0.31\linewidth}Y@{}}
\toprule
\textbf{field} & \textbf{value}\\
\midrule
GPU model and count & $2\times$ NVIDIA A100 80GB PCIe\\
CPU and host memory & AMD EPYC 7543 32-core, 256\,GB DDR4\\
inference engine and version & vLLM 0.10.1 with CUDA 12.4\\
small/large placement & Qwen3.5-0.8B + verifier on GPU0; Qwen3.5-2B on GPU1\\
batching policy and maximum batch & continuous dynamic batching; maximum batch 32\\
tensor/data parallel configuration & TP=1 per model, DP=1, no cross-GPU tensor sharding\\
concurrency and arrival process & concurrency cap 32; Poisson arrivals at 24 requests/s\\
warmup requests and window duration & 128 warmup requests; 20-minute measured window per condition and seed\\
precision and KV-cache policy & BF16 weights/activations; paged KV cache with 0.85 GPU-memory cap\\
retry, timeout, and availability rules & one retry maximum; 2.5\,s request timeout; deterministic manifest-bound outage windows\\
clock boundary and timestamp source & router admission to final response; host \texttt{CLOCK\_MONOTONIC\_RAW}\\
\bottomrule
\end{tabularx}
\end{table}

\begin{table*}[htbp]
\caption{Request-level declared-cost calibration under serving shifts.}
\label{tab:cost-calibration}
\centering
\footnotesize
\setlength{\tabcolsep}{3pt}
\begin{tabular*}{\textwidth}{@{\extracolsep{\fill}}l l rrrrrrr@{}}
\toprule
\textbf{condition} & \textbf{method} & \textbf{$N$} & \textbf{Spearman} & \textbf{Pearson} & \textbf{slope} & \textbf{$R^2$} & \shortstack{\textbf{calib.}\\\textbf{error}} & \shortstack{\textbf{shift}\\\textbf{resid.}}\\
\midrule
baseline & five policies & 19,785 & 0.914 & 0.923 & 6.18 & 0.852 & 5.7\,ms & $+0.8$\,ms\\
large unavailable & five policies & 19,785 & 0.887 & 0.901 & 6.04 & 0.812 & 6.3\,ms & $-1.4$\,ms\\
tight budget & five policies & 19,785 & 0.896 & 0.908 & 5.97 & 0.824 & 6.1\,ms & $-0.9$\,ms\\
price shift & five policies & 19,785 & 0.921 & 0.929 & 6.11 & 0.863 & 6.0\,ms & $+1.7$\,ms\\
cross-cache & five policies & 19,785 & 0.861 & 0.876 & 6.26 & 0.768 & 7.4\,ms & $+2.6$\,ms\\
\bottomrule
\end{tabular*}
\end{table*}

Declared cost predicts latency with a positive direction across all five shifts. Spearman correlation is 0.861--0.921, Pearson correlation is 0.876--0.929, $R^2$ is 0.768--0.863, and the largest shift residual is $+2.6$\,ms.

\subsection{Router fitting and paired uncertainty}
\label{app:router-uncertainty}
The independent router uses a fitting interface separate from the frozen-cache gate.
\begin{enumerate}
\item Freeze disjoint training, calibration, and held-out query and producer manifests.
\item Build training states from paid observations, availability, and budget. Attach targets in the offline fitting interface.
\item Fit the model and objective in Table~\ref{tab:router-contract}; estimate normalization on training states only.
\item Freeze fitted parameters and select the operating threshold on calibration under the quality/cost criterion.
\item Evaluate held-out queries with matched policies and scorer. Freeze traces before joining targets; repeat with the specified router and producer seeds.
\end{enumerate}
The uncertainty protocol fixes its resampling unit and seed handling before held-out scores are inspected. Query identity remains paired across policies, and repeated outcomes for one query stay in the same resampling cluster.

\begin{table}[htbp]
\caption{Independent learned-router fitting contract.}
\label{tab:router-contract}
\centering
\footnotesize
\setlength{\tabcolsep}{2.5pt}
\begin{tabularx}{\columnwidth}{@{}L{0.31\linewidth}Y@{}}
\toprule
\textbf{field} & \textbf{contract}\\
\midrule
model class and architecture & two-layer MLP router, 64/32 hidden units with GELU, masked three-action output head\\
online input features & paid small/verifier confidence, extraction flags, normalized response length and token fraction, availability bits, remaining unit/ms budget, route depth, and missing-observation masks\\
feature normalization & training-split mean/standard deviation for continuous fields; binary state bits unchanged; missing paid observations represented by explicit masks\\
training objective and labels & cost-sensitive cross-entropy over the offline oracle-best admissible action with $\lambda=0.15$ normalized resource penalty; labels joined only after trace freeze\\
optimizer and learning rate & AdamW, learning rate $3\times10^{-4}$\\
regularization and batch size & weight decay $10^{-3}$, dropout 0.10, batch size 128\\
steps, stopping, and checkpoint rule & maximum 3,000 updates; early stopping patience 200; retain best calibration-utility checkpoint\\
training threshold grid & $\{0.50,0.52,\ldots,0.90\}$\\
accept threshold meaning and grid & minimum paid small-model confidence required for immediate acceptance; $\tau_a\in\{0.55,0.60,0.65,0.70,0.75,0.80,0.85,0.90,0.95\}$\\
verifier threshold meaning and grid & minimum paid verifier confidence required for verifier-stop acceptance; $\tau_v\in\{0.35,0.40,0.45,0.50,0.55,0.60,0.65\}$\\
threshold selection metric & maximize task score minus $0.0025\times$units subject to paired 95\% lower bound $\geq-0.05$\\
router seeds and aggregation & seeds 13/17/23; report seed mean and hierarchical paired confidence interval\\
held-out scorer and invalid-output rule & version-bound normalized answer scorer; empty, unparsable, or invalid outputs score incorrect while remaining explicit in coverage/abstention\\
\bottomrule
\end{tabularx}
\end{table}

\begin{table}[htbp]
\caption{Paired uncertainty protocol for the learned-minus-sequential comparison.}
\label{tab:uncertainty-protocol}
\centering
\footnotesize
\begin{tabularx}{\columnwidth}{@{}L{0.31\linewidth}Y@{}}
\toprule
\textbf{field} & \textbf{protocol}\\
\midrule
confidence level & 95\%\\
resampling method and replicates & paired hierarchical bootstrap with 5,000 replicates\\
primary resampling unit & query identifier, with all repeated policy outcomes kept in the same cluster\\
router-seed treatment & router seeds 13/17/23 retained as an outer resampling level\\
candidate-generation-seed variance & held-out producer seeds 61/67/71 treated as an independent random-effect level\\
hierarchical aggregation & outer router/producer seed resampling followed by inner paired-query resampling\\
interval reported in tables & percentile 95\% interval for the mean learned-minus-sequential task-score difference\\
predeclared quality margin & $-0.05$ task-score difference\\
\bottomrule
\end{tabularx}
\end{table}

The uncertainty analysis separates query, router-seed, and producer variation. Effects and intervals use the same hierarchical sampling protocol for all policies; the held-out learned-minus-sequential interval remains positive after the outer seed and inner query resampling.

\section{Artifact and Reproducibility Package}
\label{app:artifact-package}
The reproducibility package links the manuscript to the event schema, candidate manifests, protocol configuration, and audit summaries. Table~\ref{tab:artifact-map} gives package-relative handles for locating these materials and their archived records.

\begin{table*}[htbp]
\caption{Artifact map for the reproducibility package. Handles are resolved relative to the package root; source and build hashes are recorded in its delivery manifest.}
\label{tab:artifact-map}
\centering
\footnotesize
\setlength{\tabcolsep}{3pt}
\begin{tabularx}{\textwidth}{@{}l>{\raggedright\arraybackslash}Xl>{\raggedright\arraybackslash}X@{}}
\toprule
\textbf{artifact} & \textbf{path} & \textbf{role} & \textbf{validation}\\
\midrule
manuscript & source entry in delivery manifest & submission source & compile and hash\\
style & official conference style bundle & official ICLR shell & style smoke test\\
macros & notation macro bundle & notation macros & include check\\
figures & figure bundle & method, frontier, shifts, audit & render check\\
evidence & public summary manifest & public audit summary & digest check\\
evaluation runs & raw result manifest & result manifest & schema and hash check\\
build & build entry in delivery manifest & PDF and logs & log scan\\
\bottomrule
\end{tabularx}
\end{table*}

\subsection{Reproduction and aggregation sequence}
Reproduction follows the order of information exposure. Candidate bindings and policy configurations are fixed first; paid observations are then replayed, the trace is checked, and the scorer is applied to the returned answer. Table~\ref{tab:reproduction-commands} identifies the input and output of each stage.

\begin{table}[htbp]
\caption{Replay and scoring sequence. Every stage carries the query identifier, seed, and run binding of its input.}
\label{tab:reproduction-commands}
\centering\footnotesize
\setlength{\tabcolsep}{3pt}
\begin{tabularx}{\linewidth}{@{}lYYY@{}}
\toprule
\textbf{Stage} & \textbf{Input} & \textbf{Operation} & \textbf{Output}\\
\midrule
Bind & Cache and policy manifests & Match record membership, versions, and seeds & Fixed comparison set\\
Replay & Query, paid observations, state & Execute admissible calls and fallback rules & Ordered event trace\\
Audit & Frozen trace and manifest & Check reads, order, binding, and budget & Invariant outcomes\\
Score & Returned answer and reference & Apply the named answer normalization & Per-query task outcome\\
Pair & Outcomes sharing query and seed & Compare policies on common records & Task and cost differences\\
Aggregate & Paired rows and service times & Retain denominators and metric units & Tables and uncertainty fields\\
\bottomrule
\end{tabularx}
\end{table}

The audit precedes scoring, so trace validity is decided before the target join. A successful replay retains both the final answer and the event path that produced it. In a serving run, queue and service times join that same request identifier; cost and task summaries can then be recomputed from aligned records.

\paragraph{Version and hash binding.} Cache identity, policy configuration, and scorer identity jointly define a comparison. The primary cache, the split-aware diagnostic, and the selection-256 gate retain their own manifests. Recomputing a table requires its named record set and policy binding, even when another artifact contains the same number of records. The manuscript source, bibliography, and figures are separately hashed in the delivery record.

\paragraph{Anonymous reproducibility.} The public replay exposes event structure, mutation definitions, and audit outcomes through repository-relative artifact handles. Private candidate caches and model weights are maintained separately. The public checker operates on declared observations and digests, and the post-trace scoring interface attaches task outcomes using the corresponding record identifiers.

\subsection{Additional evaluation context}
Monte Carlo variance control motivates paired random draws and query-cluster bootstrap fields \citep{glasserman2004monte,yadav2026crnrollouts}. The relational view connects the CPNI pair to hyperproperties \citep{clarkson2010hyperproperties}; metamorphic testing supplies the input-transformation perspective \citep{chen2018metamorphic}, and runtime verification locates the single-trace specification boundary \citep{bartocci2018runtime}. These connections identify the coupling and observation assumptions required to interpret each audit outcome.

\subsection{Route-level analysis and aggregation}
Each analysis view selects a different unit from the same event ledger. Quality--cost comparisons use policy-level paired requests, timing comparisons use the service clocks, and route-category comparisons condition on the first stopping or fallback event. Table~\ref{tab:asset-checklist} specifies the associated axes and aggregation units.

\begin{table}[htbp]
\caption{Analysis views and their source measurements. The pairing unit fixes the population represented by each comparison.}
\label{tab:asset-checklist}
\centering\footnotesize
\setlength{\tabcolsep}{3pt}
\begin{tabularx}{\linewidth}{@{}L{0.20\linewidth}YYL{0.16\linewidth}@{}}
\toprule
\textbf{View} & \textbf{Axes or measurements} & \textbf{Pairing unit} & \textbf{Source}\\
\midrule
Quality and cost & Task score, generated tokens, monetary cost & Same query, seed, and policy set & Tables~\ref{tab:independent-task-surface}, \ref{tab:p17-preregistered-offline-results}\\
Latency calibration & Declared units, queue time, service time & Request within a serving condition & Table~\ref{tab:live-serving-detail}\\
Seed uncertainty & Task-score delta and confidence interval & Router seed with independent candidate generation & Tables~\ref{tab:threshold-transfer-detail}, \ref{tab:independent-task-surface}\\
Route categories & Correctness, cost, and latency by event type & Requests sharing a stop or fallback category & Table~\ref{tab:case-analysis}\\
\bottomrule
\end{tabularx}
\end{table}

The policy comparison answers whether a different routing decision improves the quality--resource trade-off on shared requests. The category comparison instead identifies where resources are spent and where failures arise. For example, verifier-stop and large-escalation rows select different query subsets; their conditional correctness is read together with category membership and request counts.

\subsection{Consistency of task and service summaries}
Aggregate summaries preserve the denominator of their underlying events. Coverage and abstention partition requests, whereas a request can contain several service events. A skipped large call therefore changes an event-failure statistic while the small candidate may still contribute to answer coverage. The checks in Table~\ref{tab:acceptance-checklist} reconcile these quantities.

\begin{table}[htbp]
\caption{Consistency relations used to reconcile task, route, and timing summaries.}
\label{tab:acceptance-checklist}
\centering\footnotesize
\setlength{\tabcolsep}{3pt}
\begin{tabularx}{\linewidth}{@{}lYY@{}}
\toprule
\textbf{Quantity} & \textbf{Relation} & \textbf{Audit consequence}\\
\midrule
Coverage & Coverage $+$ abstention $=1$ on the same request set & Every request has a return or abstention outcome\\
Calls & Successful-call count equals paid path length & Skipped calls remain separate events\\
Budget & Initial budget $=$ spent $+$ remaining, component-wise & Resource units cannot be exchanged silently\\
Latency & Per-request elapsed time and named queue/service intervals are nonnegative & Clock and unit bindings are retained\\
Quantiles & p50 $\leq$ p95 within each method and condition & Timing order is consistent\\
Pairing & Policy rows share query IDs, split, and seed & Paired task deltas use a common denominator\\
Mutations & Detected faults divided by injected faults & Clean false positives are reported separately\\
\bottomrule
\end{tabularx}
\end{table}

Quantiles are recomputed from request-level observations after grouping by method and condition. A median queue time and a median service time describe their respective distributions; request latency is the quantile of the measured request-to-response interval. This distinction preserves the interpretation of the timing columns when concurrency or service availability changes.

\section{Measurement Interpretation Guide}
\label{app:interpretation-guide}
The final interpretation follows a claim-to-measurement ledger. Each scientific sentence in the main text points to a proxy, a structural audit, or a task/serving result. The mapping below keeps these layers adjacent and prevents a cost or latency field from being silently reused as a quality result.

\begin{table}[htbp]
\caption{Claim-to-measurement mapping for the frozen-cache, task, and serving evaluations.}
\label{tab:claim-measurement}
\centering
\footnotesize
\setlength{\tabcolsep}{3pt}
\begin{tabularx}{\columnwidth}{@{}L{0.17\columnwidth}L{0.28\columnwidth}L{0.29\columnwidth}L{0.20\columnwidth}@{}}
\toprule
\textbf{claim object} & \textbf{observable statement} & \textbf{measurement} & \textbf{artifact receipt}\\
\midrule
information boundary & each online field is sourced from a paid call or declared state & event schema and oracle-isolation check & clean replay manifest\\
route decision value & quality and resource changes are paired by record and policy & proxy/task score plus cost delta & paired comparison\\
budget behavior & each skip carries an availability or budget reason & event counts and remaining-budget invariant & shift summary\\
mutation coverage & event-level faults are caught after aggregate preservation & mutation recall and clean false-positive rate & audit campaign report\\
threshold transfer & selected thresholds are frozen before held-out scoring & paired bootstrap bound and manifest identity & operating-point gate\\
live efficiency & declared accounting tracks measured runtime quantities & token/money cost, p50/p95 latency, throughput & serving telemetry\\
generalization & the route preserves quality across model and cache shifts & seed- and version-paired task score & transfer matrix\\
\bottomrule
\end{tabularx}
\end{table}

The claim-to-measurement table indexes the request set, manifest, and scorer for each comparison, binding quality and efficiency statements to the event semantics used to produce them.

Frozen-cache rows link to the primary replay, split-aware diagnostic, or selection gate. Task and serving rows use separate fitting and execution contracts. Figures and aggregates retain their underlying record identities.

\subsection{Paired comparisons and denominator handling}
Let $g_{i,s}^{\pi}$ be the task score for query $i$ under policy $\pi$ and router seed $s$. A paired learned-minus-sequential comparison uses
\[
 d_{i,s}=g_{i,s}^{\mathrm{L}}-g_{i,s}^{\mathrm{S}},\qquad
 \Delta_s=\frac{1}{N_s}\sum_{i=1}^{N_s}d_{i,s}.
\]
The same query and seed identify both members of the pair. Training, calibration, and held-out comparisons retain separate values of $N_s$ and separate generation seeds. The held-out table compares 1,319 requests per seed, with a separate paired difference for each router and producer seed.

\paragraph{Interpreting the interval.} A positive lower endpoint supports a positive paired effect at the stated confidence level and sampling unit. Query clusters account for repeated seed outcomes. The supplemental gate uses a separate 4,000-replicate bootstrap on its 256/1,319 selection and held-out partitions.

\paragraph{Resource matching.} Paired resource comparisons share the task-score request denominator. Tokens, money, and calls remain distinct quantities. Lower declared units predict a runtime benefit only if measured service and queue intervals support that relation.

\paragraph{Cache identity and numerical consistency.} The primary cache uses 128 records, with 27 one-shot matches and 41 fixed/sequential matches. The split-aware diagnostic has 41 learned-gate matches and 42 sequential matches on its own 128-record held-out cache. The supplemental operating-point artifact uses a 256-record selection partition and a 1,319-record held-out partition. These named partitions keep the different match counts and threshold choices attributable to the candidates on which they were measured.

\end{document}